\documentclass[fleqn,usenatbib]{mnras}
\usepackage{newtxtext,newtxmath}
\usepackage[T1]{fontenc}

\DeclareRobustCommand{\VAN}[3]{#2}
\let\VANthebibliography\thebibliography
\def\thebibliography{\DeclareRobustCommand{\VAN}[3]{##3}\VANthebibliography}

\usepackage{graphicx}
\usepackage{longtable}
\usepackage{multirow}
\usepackage{amsmath}	% Advanced maths commands
\usepackage{xspace}
\newcommand\bran{AT2019dsg\xspace}
\newcommand\tywin{AT2019fdr\xspace}
\newcommand\lancel{AT2019aalc\xspace}

\usepackage{xcolor}

\title[Neutrinos from black hole flares]{Establishing accretion flares from supermassive black holes as a source of high-energy neutrinos}

\author[S. van Velzen et al.]{Sjoert van Velzen$^{1}$\thanks{E-mail: sjoert@strw.leidenuniv.nl}, 
Robert Stein$^{2,3,4}$,
Marat Gilfanov$^{5,6}$,
Marek Kowalski$^{2,3}$,
Kimitake Hayasaki$^{7,8}$,\newauthor 
Simeon Reusch$^{2,3}$,
Yuhan Yao$^{9}$,
Simone Garrappa$^{2,3}$,
Anna Franckowiak$^{10,2}$,
Suvi Gezari$^{11,12}$,\newauthor 
Jakob Nordin$^{3}$,
Christoffer Fremling$^{4}$,
Yashvi Sharma$^{9}$,
Lin Yan$^{13}$,
Erik C. Kool$^{14}$,
Daniel Stern$^{15}$, \newauthor 
Patrik M. Veres$^{10}$,
Jesper Sollerman$^{14}$,
Pavel Medvedev$^{6}$,
Rashid Sunyaev$^{6,5}$,
Eric C. Bellm$^{16}$, \newauthor 
Richard G. Dekany$^{13}$,
Dimitri A. Duev$^{4}$,
Matthew J. Graham$^{4}$,
Mansi M. Kasliwal$^{4}$,\newauthor 
Shrinivas R. Kulkarni$^{4}$,
Russ R. Laher$^{17}$,
Reed L. Riddle$^{13}$,
and Ben Rusholme$^{17}$
\\
$^{1}$Leiden Observatory, Leiden University, Postbus 9513, 2300 RA Leiden, The Netherlands \\ 
$^{2}$Deutsches Elektronen Synchrotron DESY, Platanenallee 6, D-15738 Zeuthen, Germany \\
$^{3}$Institut f\"ur Physik, Humboldt-Universit\"at zu Berlin, D-12489 Berlin, Germany \\
$^{4}$Division of Physics, Mathematics, and Astronomy, California Institute of Technology, Pasadena, CA 91125, USA \\
$^{5}$Max-Planck-Institut f\"ur Astrophysik, Karl-Schwarzschild-Str. 1, D-85741 Garching, Germany \\
$^{6}$Space Research Institute, Russian Academy of Sciences, Profsoyuznaya ul. 84/32, Moscow, 117997, Russia \\
$^{7}$Department of Astronomy and Space Science, Chungbuk National University, Cheongju 361-763,Republic of Korea \\
$^{8}$Harvard-Smithsonian Center for Astrophysics, 60 Garden Street, Cambridge, MA02138, USA \\
$^{9}$Cahill Center for Astrophysics, California Institute of Technology, MC 249-17, 1200 E California Boulevard, Pasadena, CA 91125 \\
$^{10}$Fakult\"at f\"ur Physik \& Astronomie, Ruhr-Universit\"at Bochum, D-44780 Bochum, Germany \\
$^{11}$Space Telescope Science Institute, 3700 San Martin Dr., Baltimore, MD 21218, USA \\
$^{12}$Department of Astronomy, University of Maryland, College Park, MD 20742, USA \\
$^{13}$Caltech Optical Observatories, California Institute of Technology, Pasadena, CA 91125, USA \\ 
$^{14}$The Oskar Klein Centre, Department of Astronomy, Stockholm University, AlbaNova, SE-10691, Stockholm, Sweden \\
$^{15}$Jet Propulsion Laboratory, California Institute of Technology, 4800 Oak Grove Drive, Mail Stop 169-221, Pasadena, CA 91109 \\
$^{16}$DIRAC Institute, Department of Astronomy, University of Washington, 3910 15th Avenue NE, Seattle, WA 98195, USA \\
$^{17}$IPAC, California Institute of Technology, 1200 E. California Blvd, Pasadena, CA 91125, US \\
}

\pubyear{2024}
\date{}

\begin{document}
\label{firstpage}
\pagerange{\pageref{firstpage}--\pageref{lastpage}}
\maketitle

% Abstract of the paper
\begin{abstract}
The origin of cosmic high-energy neutrinos remains largely unexplained. For high-energy neutrino alerts from IceCube, a coincidence with time-variable emission has been seen for three different types of accreting black holes: (1) a gamma-ray flare from a blazar (TXS 0506+056), (2) an optical transient following a stellar tidal disruption event (TDE, AT2019dsg), and (3) an optical outburst from an active galactic nucleus (AGN, AT2019fdr). For the latter two sources, infrared follow-up observations revealed a powerful reverberation signal due to dust heated by the flare. This discovery motivates a systematic study of neutrino emission from all supermassive black hole with similar dust echoes. Because dust reprocessing is agnostic to the origin of the outburst, our work unifies TDEs and high-amplitude flares from AGN into a population that we dub accretion flares. Besides the two known events, we uncover a third flare that is coincident with a PeV-scale neutrino (AT2019aalc). Based solely on the optical and infrared properties, we estimate a significance of 3.6$\sigma$ for this association of high-energy neutrinos with three accretion flares. Our results imply that at least $\sim$10\% of the IceCube high-energy neutrino alerts could be due to accretion flares. This is surprising because the sum of the fluence of these flares is at least three orders of magnitude lower compared to the total fluence of normal AGN. It thus appears that the efficiency of high-energy neutrino production in accretion flares is increased compared to non-flaring AGN. We speculate that this can be explained by the high Eddington ratio of the flares. 
\end{abstract}


\section{Introduction} \label{sec:intro}
Accreting black holes have long been suggested as potential sources of high-energy particles \citep{fg08,murase_agn} and this expectation was supported by the detection of a high-energy neutrino coincident (at the 3$\sigma$-level) with gamma-ray flaring from the blazar TXS\,0506+056 \citep{txs_mm} and from the nearby active galactic nucleus (AGN) NGC 1068 at the 4$\sigma$-level \citep{IceCube_NGC1068_22}. There is also evidence to support neutrino emission from the broader blazar population \citep[see e.g][]{Giommi20,Plavin20,hovatta_20,kun_21, Buson22}, but blazars alone cannot account for the observed high-energy neutrino flux \citep{IC17_LAT,Aartsen:2019fau,hooper_19,luo_20}. Similar to the electromagnetic sky, we expect that the observed cosmic neutrino flux \citep{ic_astro} arises from multiple source populations \citep{pie-chart}. 
% add more citations?

Optical follow-up observations of IceCub neutrino alerts \citep{ic_realtime, IceCat-1} using the Zwicky Transient Facility (ZTF; \citealt{Bellm19,Graham19,Stein23}) have identified two optical flares from the centers of galaxies coincident with PeV-scale neutrinos: \bran \citep{Stein20} and \tywin \citep{Reusch22}. The former belongs to the class of spectroscopically-classified tidal disruption events (TDEs) from quiescent black holes, while the latter originated from a type~1 (i.e., unobscured) AGN (though see \citealt{pitik_22} for an alternative interpretation). The distinctive shared properties we present below suggest these two flares could share a common origin.

Both events show a large amplitude optical flare with a rapid rise time, signalling a sudden increase of mass accretion rate onto the supermassive black hole. Of the $\sim 10^4$ AGN detected by ZTF \citep{vanVelzen20}, less than 1\% show similarly rapid and large outbursts \citep{Reusch22}. The most important unifying signature of the two neutrino-coincident ZTF sources is delayed transient infrared emission, detected by NEOWISE \citep{Wright10,Mainzer14}. This infrared emission is due to reprocessing of the optical to X-ray output of the flare by hot dust ($T \sim 2 \times 10^3$~K) at distances of 0.1-1~pc from the black hole \citep{Lu16,vanVelzen21_ISSI}. 

A dust reverberation signal is largely agnostic to the origin of the flare near the black hole. Any transient emission at optical, UV or X-ray wavelengths that evolves on a timescale that is shorter than the light travel time to the dust sublimation radius ($R_{s} \sim 0.1$~pc) will yield a similar-looking dust echo: a flat-topped light curve with a duration of $2 R_s/c$. This implies that infrared observations of these echoes can be used to construct a sample that unifies ``classical TDE" (such as \bran) and extreme AGN flares (such as \tywin). In this work, we collect a sample of dust echoes from nuclear flares and investigate the significance of their correlation with high-energy neutrinos\footnote{While this paper was under review, evidence for a correlation of high-energy neutrinos and a different sample of infrared-selected TDE candidates \citep{Jiang21b} was presented by \citet{Jiang23}.}. 

This paper is organized as follows. In section~\ref{sec:samples} we present the details of the two samples: extreme nuclear transients from supermassive black holes (i.e., {\it accretion flares}) and high-energy neutrinos from IceCube. In section~\ref{sec:stat} we then compute the significance of a correlation between the two samples. This statistical analysis is based only on optical and infrared data. In section~\ref{sec:mm} we include information from other wavelengths: radio, X-ray, and gamma-rays. In section~\ref{sec:impl} we discuss the implications of the results. %and we conclude with a summary. 

\section{Catalog construction}\label{sec:samples}
To build our sample of accretion flares, we select transients from the centers of galaxies as measured with ZTF data and then search for a significant infrared flux increase after the peak of the optical flare using NEOWISE observations. 

Below we first present the details of the flare selection and our estimates of the black hole mass. We then present the properties of the IceCube neutrino sample and our definition for a flare-neutrino coincidence.  

\subsection{ZTF nuclear flare selection}\label{sec:ztf}
\begin{figure}
  \centering
  \includegraphics[width=0.5\textwidth]{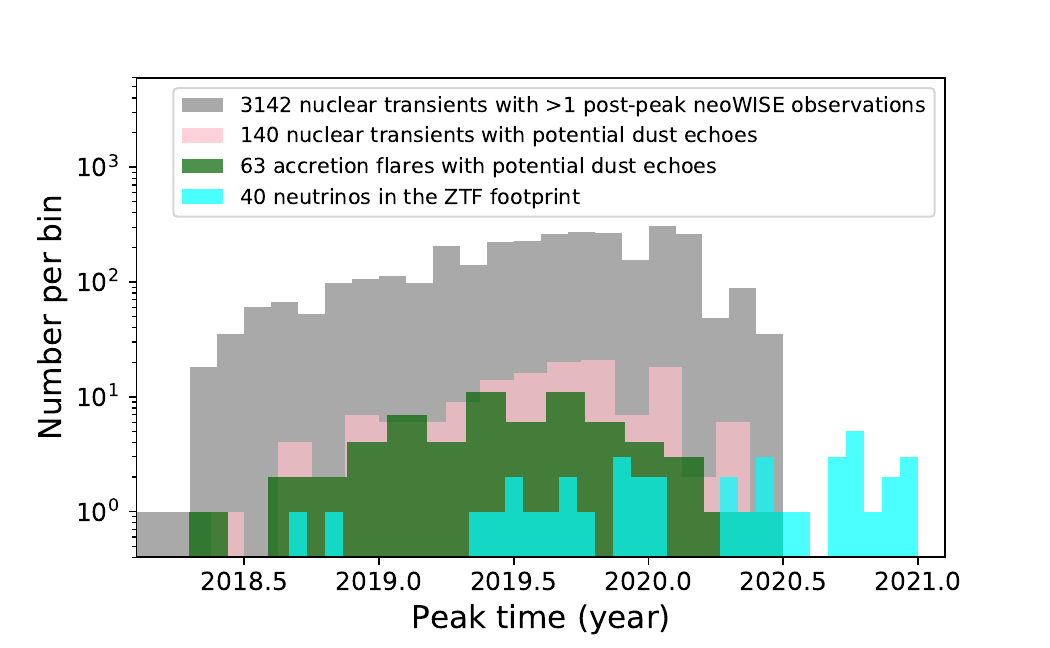}
  \caption{Distribution of the time of maximum light of the ZTF light curve and the detection date of the IceCube alerts. The lack of ZTF events with a peak after mid-2020 happens because the NEOWISE data release includes observations up to the end of 2020 and to be able to measure the properties of the dust echo we require at least two post-peak detections in NEOWISE. Given the 6 month cadence of NEOWISE, the ZTF light curve has to peak prior to mid-2020 to meet this requirement. Because we allow the neutrino to arrive after the peak of the optical flare, all IceCube events up to the end of 2020 are included in the analysis. }
  \label{fig:years}
\end{figure}

\begin{figure}
  \centering
  \includegraphics[width=0.5\textwidth]{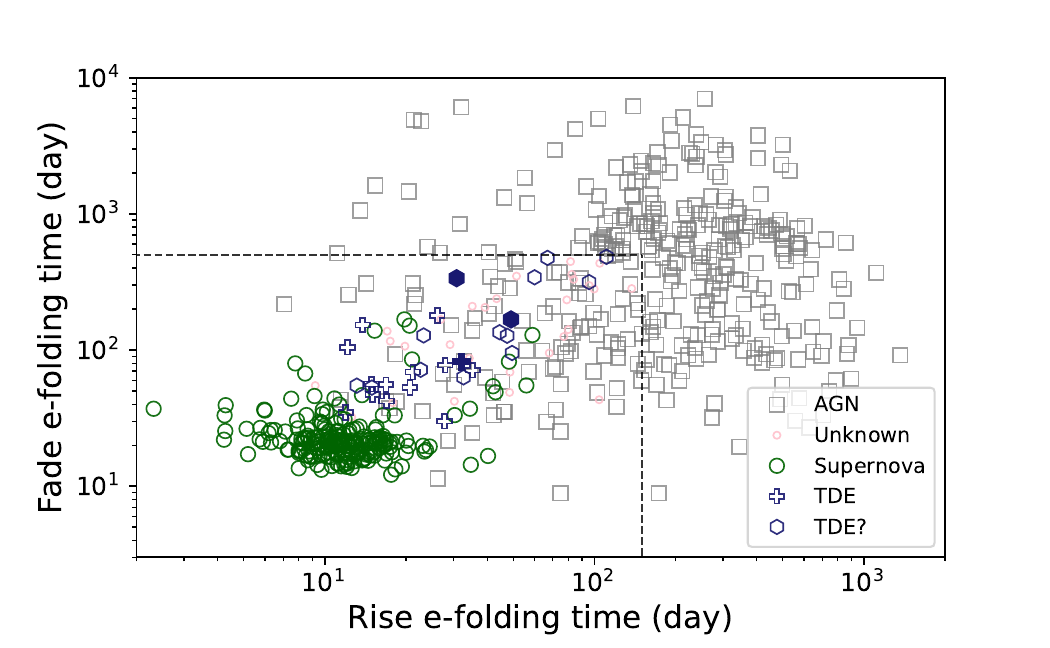}
  \includegraphics[width=0.5\textwidth]{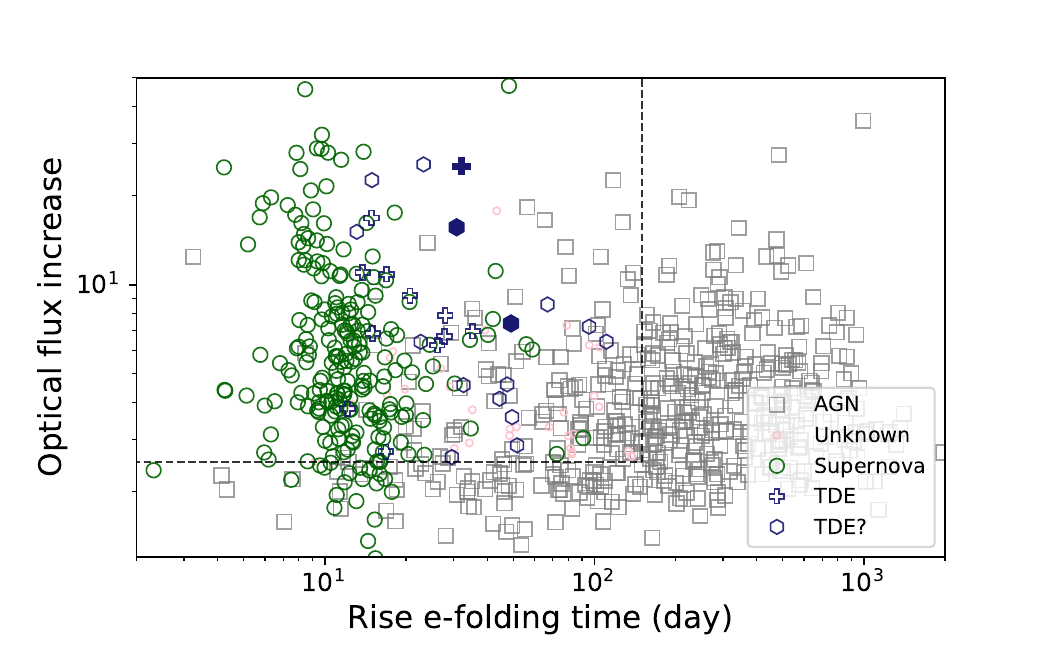}
  \caption{Parameters inferred from a Gaussian rise plus exponential decay model applied to the ZTF light curves. The flux increase is measured relative to the ZTF reference image. The dashed lines indicates the box that is used to separate accretion flares from regular AGN variability. This requirement selects nuclear supernovae plus all spectroscopically-confirmed ZTF TDEs. The label `TDE?' indicates accretion flares that occurred in active galaxies (i.e., sources with evidence for accretion prior to the main flare). The three events coincident with a high-energy neutrino are indicated with solid symbols.}
  \label{fig:risefade}
\end{figure}

%The parent sample of the accretion flares with dust echoes to be used in the correlation analysis is constructed from ZTF \citep{Bellm19,Graham19,Dekany20} and NEOWISE \citep{Mainzer11,WISE} data. 

%We start by selecting ZTF alerts \cite{Masci09} that are consistent with originating from the center of a galaxy. 
As described in \citet{vanVelzen18_NedStark,vanVelzen20} processing of the ZTF alert stream \citep{Masci19,Patterson19} to yield a sample of nuclear transients is done with AMPEL \citep{Nordin19}. The input streams include both public ZTF data (MSIP) and private partnership data. 
We remove events with a weighted host-flare offset \citep{vanVelzen18_NedStark} $>0.5"$. To be able to measure the properties of the light curve we require at least 10 ZTF detections. We also remove ZTF sources for which the majority of the light curve measurements have a negative flux relative to the reference image. These requirements leaves 3142 nuclear transients, see Fig.~\ref{fig:years}.  

To measure the peak flux of the ZTF light curve, we use the observation with the highest flux, after restricting to 90\% of the data points with the highest signal-to-noise ratio (excluding 10\% of lower quality data makes the peak estimate robust against outliers that occasionally occur in ZTF data). 

To define our sample of accretion flares we use a requirement on the rise-timescale and fade-timescale of the ZTF light curve, see Fig.~\ref{fig:risefade}. These timescales are obtained from the measurements of \citet{vanVelzen20} who applied a Gaussian-rise exponential-decay model to the ZTF alert photometry (both the $g$- and $r$-band are used in this fit). This model explicitly assumes that a single transient explains the entire ZTF light curve. When a light curve has multiple peaks of similar amplitude, the parameters of the fit reflect the (slower) timescale of the majority of the data points.
To remove regular variability from normal AGN, we require a minimal amplitude of the flare of $\Delta m<-1$  ($m$ denotes the magnitude) and also we set an upper limit to the rise and fade timescale (e-folding time $<150$~days and $<500$~days, respectively). These cuts leave  1732 sources. The cuts on amplitude and rise/fade times are designed to cast a wide net, recovering all ZTF TDEs and all large-amplitude flares from Seyfert galaxies that have been reported in earlier work \citep{vanVelzen20,Frederick20,Hammerstein21}. About 15\% of the sources are spectroscopically confirmed SN.

To keep track of the follow-up resources and spectroscopic classifications we used the GROWTH Marshal \citep{Kasliwal19}. Most of the supernova classifications (e.g. as shown in Fig.~\ref{fig:risefade}) are based on SEDM \citep{Blagorodnova18,Rigault19} data obtained for the ZTF Bright Transient Survey project \citep{Fremling20,Perley20}. 

\subsection{NEOWISE dust echo selection}\label{sec:neoWISE}
\begin{figure}
  \centering
  \includegraphics[width=0.5\textwidth]{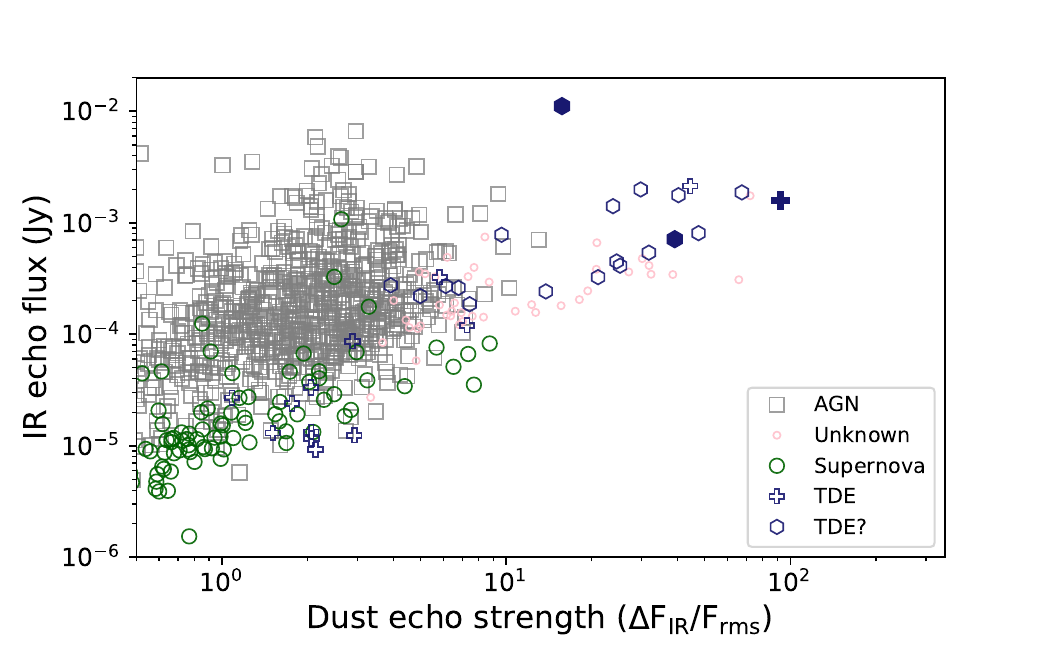}
  \includegraphics[width=0.5\textwidth]{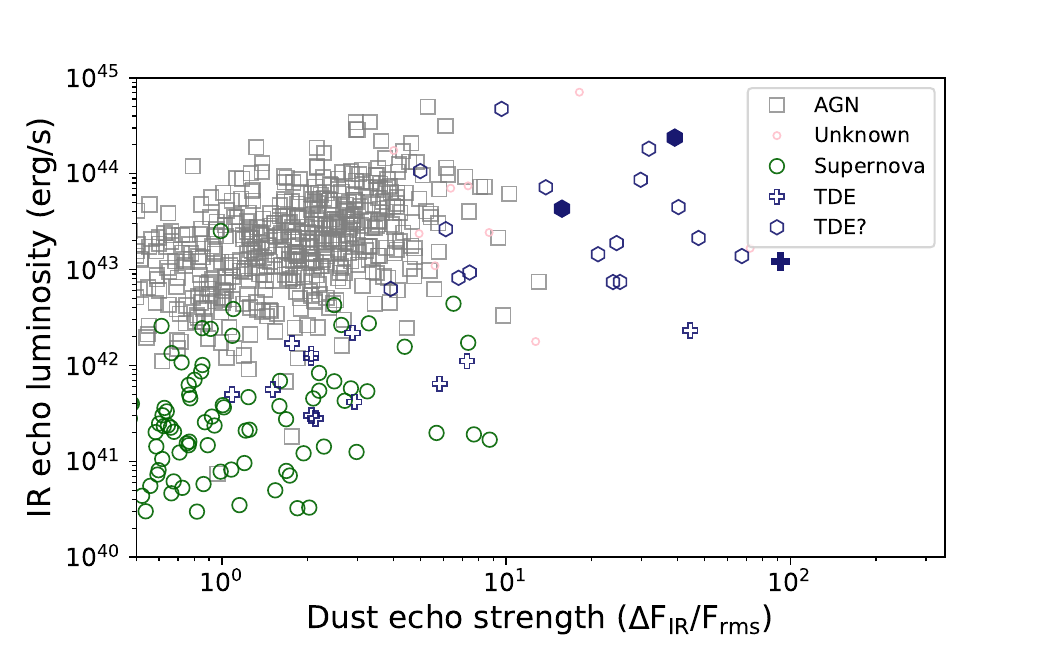}
  \caption{The dust echo flux and luminosity of nuclear transients in ZTF. We see that the three accretion flares coincident with a high-energy neutrino (filled symbols) are among the strongest  dust echoes in the sample of nuclear transients. }
  \label{fig:df}
\end{figure}

The NEOWISE \citep{Mainzer14} light curves cover a period from 2016 to 2020. Most parts of the sky are visited every 6 months and receive about 10 observations within a 24 hour period of this visit \citep{Wright10}. The inverse-variance weighted mean of the cataloged flux during these visits is used to construct the NEOWISE light curves. For each source, the baseline is defined using all NEOWISE observations obtained up to 6 months before the peak of the ZTF light curve. The 6 months padding is added to avoid including part of the dust echo signal into the baseline (e.g., when ZTF observations miss the onset of the flare). 

To measure the echo strength we require two observations after the ZTF peak. We define the dust echo flux as $\Delta F_{\rm IR}$: the difference between the baseline flux and the mean NEOWISE flux within one year after the ZTF peak. The echo strength is $\Delta F_{\rm IR}/F_{\rm rms}$, where $F_{\rm rms}$ is the root-mean-square variability of the baseline observations. The significance of the rms variability is measured using the ratio $F_{\rm rms}/\sigma_F$, where $\sigma_F$ is the measurement uncertainty of the baseline observations. We selected candidate dust echoes by requiring that the echo strength is larger than the significance of the baseline rms variability: $\Delta F_{\rm IR}/F_{\rm rms}>F_{\rm rms}/\sigma_F$. We apply this criterion to the light curves of both NEOWISE bands (W1 and W2; central wavelengths of 3.4 and 4.6 $\mu$m, respectively). This selection leaves 140 nuclear transients with candidate dust echoes. After selecting accretion flares based on the ZTF properties (Fig.~\ref{fig:risefade}), we are left with 63 flares with candidate dust echoes (Table~\ref{tab:ZW}). In Fig.~\ref{fig:df} we show the echo flux and luminosity (for the subset of sources with a spectroscopic redshift) versus echo strength.

The time difference between the optical and infrared light curve peaks yields an estimate of the inner radius of the dust reprocessing region. At this dust sublimation radius, the bolometric flux absorbed by the dust is equal to the infrared luminosity emitted by the dust (with a spectrum that is determined by the sublimation temperature of the dust). We can therefore estimate the bolometric luminosity of the flare from the duration of the infrared reverberation light curve \citep{Lu16, vanVelzen16b,vanVelzen21_ISSI}. Using this geometric luminosity estimate (Eq.~12 in \citealt{vanVelzen16b}), we find a bolometric luminosity $\sim 10^{45}\,{\rm erg}\,{\rm s}^{-1}$ for all accretion flares coincident with high-energy neutrinos. All are consistent with reaching the Eddington limit (Table~\ref{tab:mm}). 
For infrared emission due to reverberation, the energy emitted by the dust cannot exceed the integrated bolometric luminosity of the flare. For this reason, the lower optical-to-infrared ratio of the third source (cf. Fig.~\ref{fig:lcs}) likely implies a larger bolometric correction for its optical emission. This suggests $\approx 1$~mag of optical extinction. 

\begin{figure*} % Figure by Yuhan
  \centering
  \includegraphics[width=0.8\textwidth]{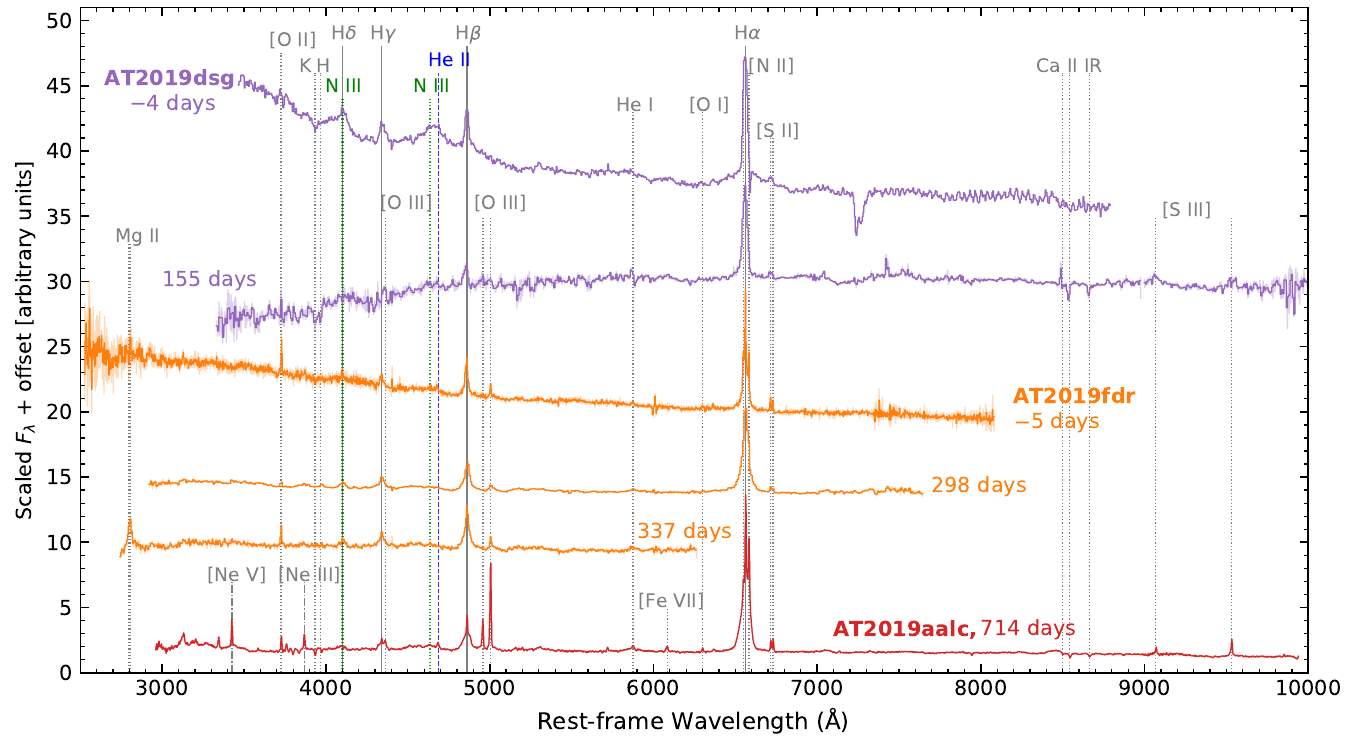}
  \caption{Optical spectra of the three accretion flares coincident with a high-energy neutrino. The observed time relative to the peak of the optical emission is indicated. Including data from \citet{Frederick20,Stein20,Reusch22}.}
  \label{fig:spectra}
\end{figure*}

\subsection{Black hole mass estimates}\label{sec:mbh}
For all nuclear flares in our sample, black hole masses are can be estimate if optical spectra are available. We either use a relation based on reverberation mapping \citep{Blandford82,Peterson04}, also known as ``virial mass estimates", or the $M$--$\sigma$ relation \citep{Magorrian98,Gebhardt00}. We use the former for spectra of type~1 AGN, i.e., sources that show broad Balmer emission lines in their optical spectrum. The $M$--$\sigma$ relation is used for sources without broad emission lines that have a host galaxy spectrum with a measurement of the velocity dispersion of the stars. 

For the reverberation method, a measurement of the size of the broad-line region in combination with the velocity of the emission lines yields an estimate of the black hole mass. This distance of the broad-line region to the black hole is not measured directly, but follows from the observed disk luminosity. We adopt the relation from \citet{Ho15}:
\begin{eqnarray}
    \log(M_{\rm BH} / M_\odot) &=& \log \Bigg[\left(\frac{{\rm FWHM~(H}\beta)}{\rm 1000~km~s^{-1}}\right)^2 \times \nonumber \\ 
    && \left(\frac{L_{5100}}{10^{44}\,{\rm erg}\,{\rm s}^{-1}}\right)^{0.533} \Bigg] + 6.91  \label{eq:virial}
\end{eqnarray}
Here $L_{5100}$ is the continuum luminosity at 5100 \AA\ in the rest frame. 

For active galaxies with spectra from the Sloan Digital Sky Survey (SDSS; \citealt{york02}) we use the catalog of \citet{Liu19_SDSS}, who selected 14,584 type 1 AGN based on detection of a broad H$\alpha$ line and applied Eq.~\ref{eq:virial} to measure the black hole mass. We find 580 nuclear transients with black hole mass estimates based on this catalog. Of these, eight are classified as accretion flares with potential dust echoes. In addition, spectroscopic follow-up observations of ZTF transients have yielded nine more type~1 AGN, six of which have been published \citep{Frederick19,Frederick20} and three are presented for the first time in this work: AT2019meh (ZTF19abclykm), AT2020afab (ZTF19abkdlkl), and AT2020iq (ZTF20aabcemq). We also obtained a new post-peak spectrum of \lancel (Fig.~\ref{fig:spectra}), which shows ongoing accretion two years after the peak of the optical flare. 

The spectra of AT2020iq, AT2019meh, and \lancel were obtained with the Low Resolution Imaging Spectrograph (LRIS; \citealt{Oke95}) on the 10-m Keck-I telescope at 20, 660, and 714\,days post $t_{\rm peak}$ (Table~\ref{tab:ZW}), respectively. The new spectrum of AT2020afab was obtained with the Double Spectrograph (DBSP; \citealt{Oke82}) on the 5-m Palomar telescope (P200) at 15\,days post $t_{\rm peak}$.
The DBSP spectrum was reduced with the \texttt{pyraf-dbsp} pipeline \citep{BellmDBSP}. The LRIS spectra were reduced using \texttt{Lpipe} \citep{PerleyLpipe}.

For the remaining transients that have a velocity dispersion measurement based on a spectrum of the host galaxy, we apply the \citet{Gultekin09} $M$--$\sigma$ relation. This yields 219 additional $M_{\rm BH}$ measurements, of which five are classified as accretion flares with potential dust echoes. Of these five, three are based on archival SDSS spectra of the host galaxy \citep{Thomas13} and two are based on follow-up observations obtained after the flare has faded \citep{Nicholl20,Cannizzaro21}. 
%Finally, for sources with a measured redshift, but without a measured velocity dispersion, we estimate the black hole mass from the ZTF $r$-band luminosity of the host galaxy, using the correlation between $r$-band bulge luminosity and black hole mass \citep{Tundo07}. % not applied, but would have been useful 
In Table~\ref{tab:ZW}, we list the reference for the black hole mass estimate for each accretion flare. 

\subsection{IceCube neutrino alerts}\label{sec:IC}
Our parent sample of neutrino alerts \citep{ic_realtime} includes the events published up to December 31, 2020. We exclude the IceCube alerts that were subsequently retracted after the automated alert was issued. We also remove two events without a reported signalness (IC190331A; \citealt{ic190331a} and IC200107A;  \citealt{ic200107a}) and one event  with a 90\%CL area larger than 300~deg$^2$ (IC200410A;  \citealt{ic200410a}). This leaves 43 events, listed in Table~\ref{tab:IC}. Of these, three fall outside ZTF extra-galactic footprint ($\Omega_{\rm ZTF}=2.8\times 10^4$~deg$^2$; \citealt{Stein20}), leaving 40 IceCube alerts that can yield a coincidence with an accretion flare in our ZTF+NEOWISE dataset. 

The signalness, or $P_{\rm astro}$, measures the probability that a detector track recovered by IceCube is from a cosmic neutrino, based on the reconstructed energy and assumptions about the astrophysical neutrino flux. 
The sum of signalness of the 40 neutrinos in the ZTF footprint is 17.7. 
Multiplying by 0.9 to account for the 10\% of neutrinos  whose true location is outside the reported 90\%CL area, we estimate that about 16 cosmic neutrinos can in principle be recovered by our analysis.

%To arrive at the expected number of cosmic neutrinos that can be recovered by our analysis we should correct for the fact that neutrinos from late-2020 have a lower probability to be coincident with an accretion flare. Since current NEOWISE data release includes observations up to the end of 2020 and we require two NEOWISE detections to establish a potential dust echo, neutrinos that arrive after mid-2020 can found in coincidence only for a fraction of the 1 year window of our search (see Fig.~\ref{fig:years}). After taking this reduced detection window into account, the total signalness is 16.5. 

\begin{figure}
  \centering
  \includegraphics[width=0.5\textwidth]{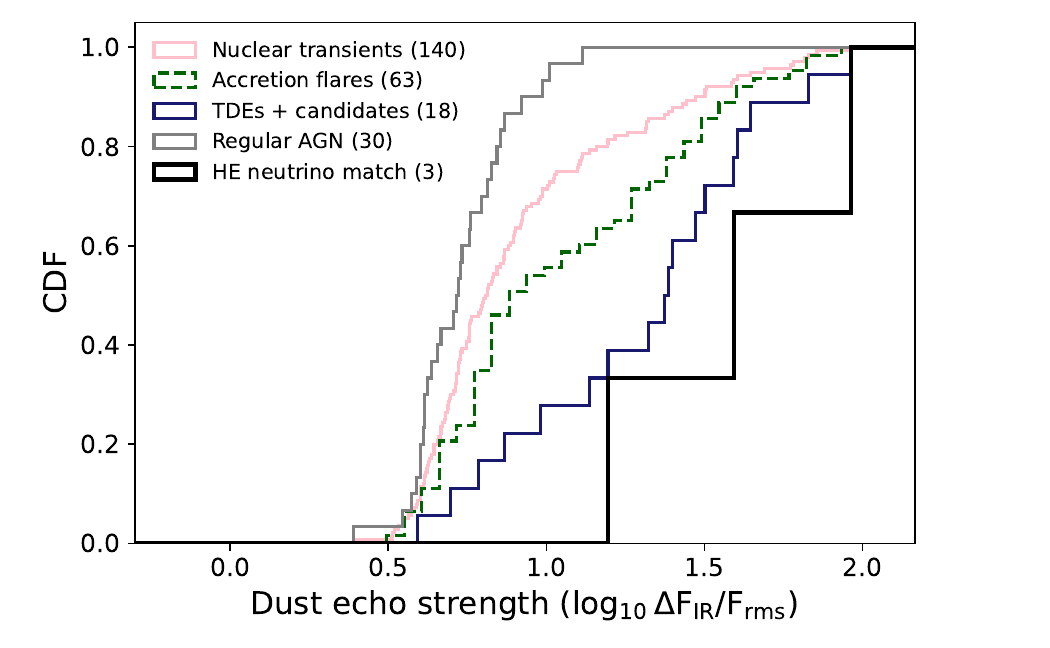}
  \caption{Flares coincident with high-energy neutrinos have very strong dust echoes. 
  The TDE candidates (blue line) are defined as accretion flares with an estimated black hole mass (Fig.~\ref{fig:df_mbh}) below $10^{8}\,M_\odot$. This population shows much stronger dust echoes compared to regular variability from the AGN population (grey line). 
  In the full population of 140 nuclear transients with candidate dust echoes (pink line) we find three events coincident with a high-energy neutrino. The echo strength of these three flares is inconsistent with originating from the full population ($p=0.007$ for an Anderson-Darling test). }
  \label{fig:cum}
\end{figure}

\subsection{Flare-neutrino coincidence}
An accretion flare is considered coincident with a neutrino when the source falls inside the 90\%CL reconstructed neutrino sky location and this flare is detected in ZTF and NEOWISE when the neutrino arrives, with a maximum delay of one year relative the peak of the optical light curve. For longer delays, our search would lose sensitivity because our NEOWISE dataset only contains photometry up to the end of 2020 (see Fig.~\ref{fig:years}). This requirement for coincidence yields three matches between the 43 IceCube alerts (section~\ref{sec:IC}) and the 63 accretion flares with potential dust echoes (section~\ref{sec:neoWISE}). 

We immediately notice that the three accretion flares with a coincident neutrino have very strong dust dust echoes compared to the rest of the nuclear flare population (\ref{fig:cum}). The significance of this result is discussed below in section~\ref{sec:stat})

\begin{figure}
  \centering
  \includegraphics[width=0.5\textwidth]{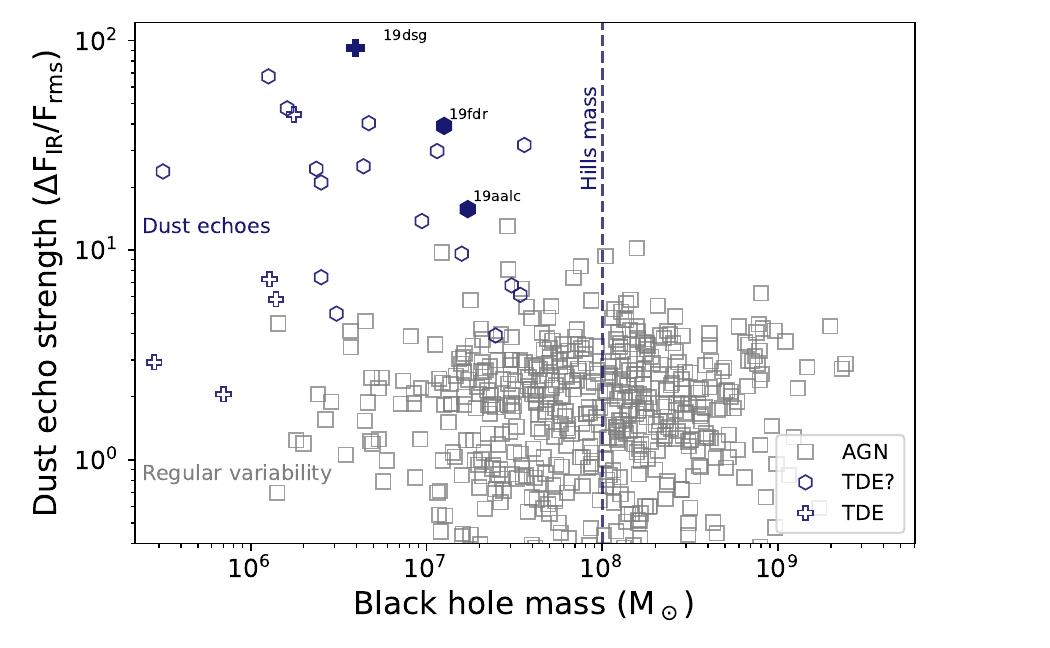}
  \caption{Significant dust echoes occur almost exclusively in low-mass black holes. The onset of strong echoes, measured using the infrared flux increase within one year of the optical peak of the flare, coincides with the Hills mass \citep{Hills75} for a solar-type star (defined by the requirement that the tidal radius is larger than the black hole horizon). The label `TDE?' indicates accretion flares that occurred in AGN. The three accretion flares coincident with a high-energy neutrino are indicated with filled symbols. }
  \label{fig:df_mbh}
\end{figure}

Black hole mass estimates based on optical spectroscopy are available for about 1/3 of the accretion flares (see section~\ref{sec:mbh}). We find that strong dust echoes ($\Delta F_{\rm IR}/F_{\rm rms}>10$) are observed almost exclusively for events with $M_{\rm BH}<10^{8}\,M_\odot$, see Fig.~\ref{fig:df_mbh}. The threshold for strong echoes appears to coincide with the maximum mass for a visible disruption from a solar-type star \citep{Hills75}. This motivates the construction of a {TDE candidate} sample: all accretion flares with $M_{\rm BH}<10^{8}\,M_\odot$. The infrared properties of the accretion flares with black hole mass estimates are used to place all nuclear transients with dust echoes into a two-tier classification scheme: (1) accretion flares with strong dust echoes and (2) regular AGN variability. In the next section, we consider the hypothesis that the former class is a source of high-energy neutrinos.

\section{Statistical significance}\label{sec:stat}
We measure the strength of the observed flare-neutrino correlation by defining a test statistic based on the likelihood ratio. Our statistical test accounts for both the spatial localization of the neutrino and the infrared properties of the flare relative to the TDE candidate population. To quantify the statistical significance of our result we randomly redistribute the accretion flares into the ZTF footprint and compute the test statistic for a large number of these simulations. %We measure a probability of $1.5\times 10^{-4}$ to obtain the observed test statistic by chance, implying a significance of 3.6$\sigma$ for the three observed neutrino coincidences.

Similar to the approach used for the neutrino detection from the blazar TXS\,0506+056 \citep{txs_mm}, our likelihood analysis is not blind because the input was defined after two neutrino associations had been established (\bran and \tywin). However, we note that these prior associations were based on ZTF follow-up observations of neutrino alerts. The infrared data relevant for our analysis were released in March 2021, and therefore the echo strength was not used to establish the neutrino coincidence. %Yet \bran has the strongest dust echo of all ZTF flares (Fig.~\ref{fig:cum}). Thanks to this remarkable property, the parameter choices of our statistical method are straightforward and have only a limited effect on the final outcome. We confirm this by computing the probability of a change coincidence using solely the low areal density of strong echoes (see Supplementary Materials). 

Below we present the details of our likelihood ratio test (section~\ref{sec:stattest}), including its statistical power (section~\ref{sec:power}) and robustness (section~\ref{sec:robust}) we also demonstrate how the significance can be estimated without the use of Monte Carlo simulations (section~\ref{sec:dens}).  The software and the data required to reproduce these likelihood estimates can be obtained via Zenodo \citep{HelloLancel}.

\subsection{Likelihood-ratio based estimate of significance}\label{sec:stattest}
To test for a correlation between high-energy neutrinos and accretion flares we consider the likelihood ratio between a signal hypothesis and a background hypothesis. 
For a given neutrino, $\nu$, there are are several possible hypotheses that can be used to explain the origin. If there are $N$ accretion flares, we can test $N$ discrete hypotheses, i.e., that the neutrino originated from source \#1 ($H_{1}$), from source \#2 ($H_{2}$), etc. We can denote this group of hypotheses as signal hypotheses, i.e., $H_{i} \equiv S_{i}$. Additionally, we have a further possible explanation, namely that the neutrino did not originate from any of the accretion flares. The neutrino itself might not be of astrophysical origin (but instead is due to the atmospheric background), or the neutrino may originate from an astrophysical source that is not included in our sample of accretion flares. All these alternative options are included in the hypothesis $H_{0}$. We can also denote this null hypothesis, or background hypothesis, as $H_{0} \equiv B$.

For each hypothesis, we can define the likelihood that the data, $x$, is obtained under that hypothesis $\mathcal{L}_{i} = P(x | H_{i})$. For each neutrino, we select the hypothesis with the greatest likelihood as our best fit:

\begin{equation}
    \hat{\mathcal{L}} = \max\left( \mathcal{L}_{i} \right)
\end{equation}

In the rare occasion that multiple accretion flares are found in coincidence with a single neutrino, we select the flare with the highest likelihood:
\begin{equation}
    \hat{\mathcal{L}_{s}} = {\rm max} \left( \mathcal{L}_{i>0} \right)
\end{equation}
Next, we compare the signal hypothesis to the background hypothesis: 
\begin{equation}
    \hat{\mathcal{L}} = {\rm max} \left(\hat{\mathcal{L}_{s}}, \mathcal{L}_{0} \right)
\end{equation}
When no accretion flare is found in coincidence with a neutrino, we have $ \hat{\mathcal{L}} =\mathcal{L}_0$. 

Below, we first discuss the input for the signal hypothesis, followed by the components of the background hypothesis.

\subsubsection{Components of the signal hypothesis}
The probability of the data under a particular signal hypothesis is given by the joint probability of two distinct components:
\begin{enumerate}
    \item  The probability that a spatial and temporal coincidence between a given neutrino alert and accretion flare would be observed if the neutrino was produced by the accretion flare. We denote this with $P_{\rm coin}(\alpha,\delta, t|S)$, with $\alpha$ and $\delta$ the right ascension and declination of the accretion flare, respectively. 

    \item The probability that the properties of the accretion flare would be observed, if the neutrino was produced by the accretion flare. 
\end{enumerate}

Probability 1 of this list is trivial to calculate, because spatial coincidence is established when the accretion flare falls inside the reported 90\%CL localization area of the neutrino. Hence for coincident events, $P_{\rm coin} = 0.9$. Here we implicitly assume that all neutrinos emitted by accretion flares arrive within our temporal search window. %This term can be though of as a simplified point-spread-function (PSF) of the observation. The information required to construct the actual spatial PSF of each IceCube alert is not made public. 

Probability 2 of the list yields a second and third term that enter the likelihood of the signal hypothesis. These are based on the echo strength and the echo flux. We motivate these two terms below.

\begin{figure}
  \centering
  \includegraphics[width=0.5\textwidth]{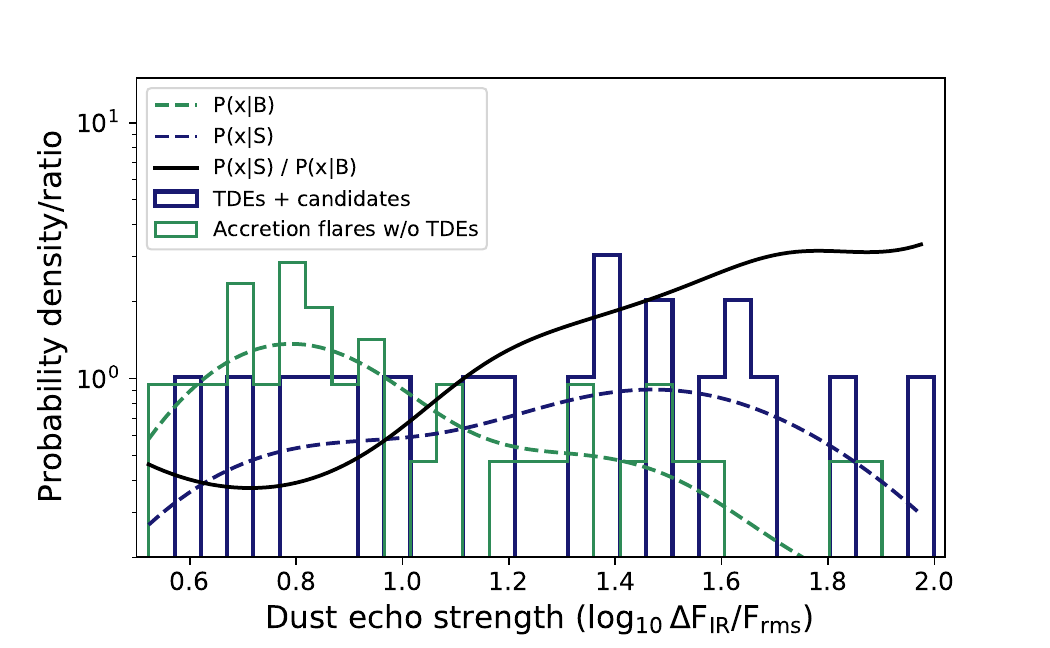}\
  \caption{The observed distribution of dust echo strength for the full population of nuclear transients with variable infrared emission (green). The observed distributions are approximated using using kernel density estimation (dotted lines). The ratio of these two probability density functions (solid black line) is used in the likelihood analysis (Eq.~\ref{eq:TS})}\label{fig:h_strength}
\end{figure}

\begin{figure}
  \centering
  \includegraphics[width=0.5\textwidth]{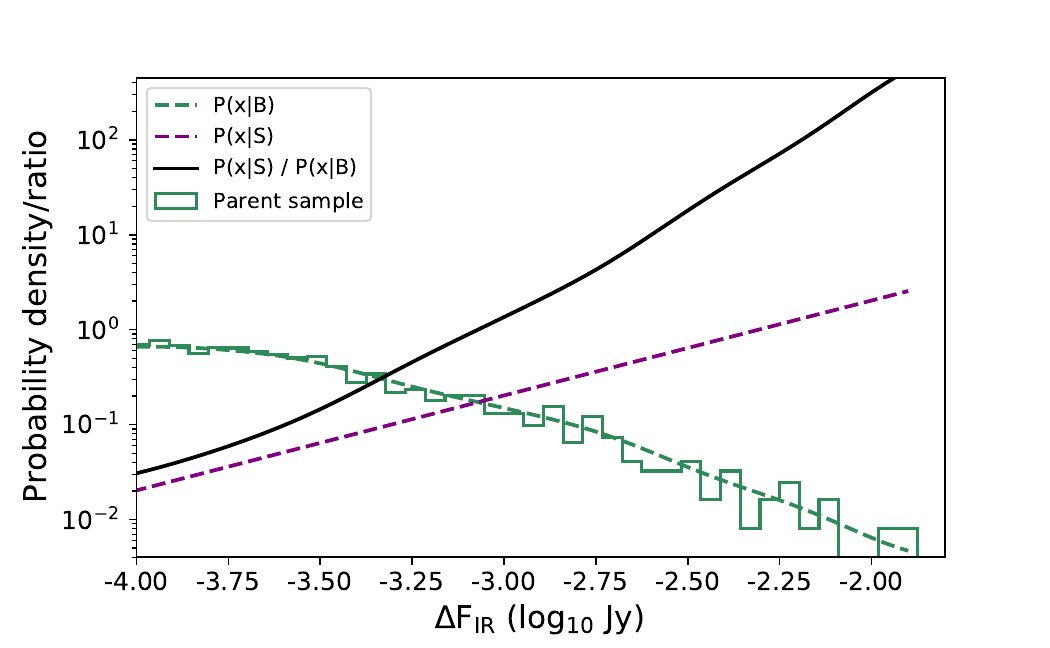}
  \caption{The observed distribution of echo flux for the population of nuclear transients with post-peak NEOWISE detections (green). From this parent sample, a PDF is obtained using kernel density estimation (dotted lines). For the signal hypothesis we assume $P(\rm \Delta F_{\rm IR}|S)\propto \Delta F_{\rm IR}$ (purple line).  The observed echo flux of the three events that found in coincidence with an IceCube neutrino alert is consistent with this PDF (See Fig.~\ref{fig:h_flux_predict}).}
  \label{fig:h_flux}
\end{figure}

A key property of our signal hypothesis is that stronger echoes are less likely to be explained by normal AGN variability (Fig.~\ref{fig:df_mbh}), but rather by a single outburst or transient. We estimate the PDF of the echo strength from the observed distribution of the 18 accretion flares that are candidate TDEs (i.e., sources with an estimate black hole mass $<10^8\,M_\odot$). To turn the binned distribution into a continuous PDF, we use a Gaussian kernel density estimate (KDE). The result is shown in Fig.~\ref{fig:h_strength}. For all KDE estimates, we select the optimal bandwidth following Scott’s Rule \citep{Scott15}. 

The use of echo flux in the signal likelihood is motivated by applying what is arguably the simplest-possible model for neutrino production: a linear coupling between the total electromagnetic luminosity and the neutrino luminosity. To estimate the neutrino flux at Earth, we thus need the total electromagnetic fluence (i.e., the bolometric energy over the distance squared). The bolometric energy cannot be measured directly because a large (but unknown) fraction of the energy is emitted at higher frequencies than the optical wavelength range of ZTF. Fortunately, dust absorption is efficient up to soft X-ray frequencies, hence the dust reprocessing light curve provides a good tracer of the bolometric luminosity. With these simplifying assumptions, we thus obtain a linear scaling of the neutrino flux with infrared flux of the echo: 
\begin{equation}
    P(\Delta F_{\rm IR}|S) = \frac{\Delta F_{\rm IR}}{{\rm max}(\Delta F_{\rm IR})-{\rm min}(\Delta F_{\rm IR})}
\end{equation}
Here the max/min correspond to the maximum/minimum observed echo flux of the accretion flares, respectively. After applying our test statistic, we will validate the consistency of this linear scaling. 

\begin{figure*}
\centering
\includegraphics[width=0.48\textwidth]{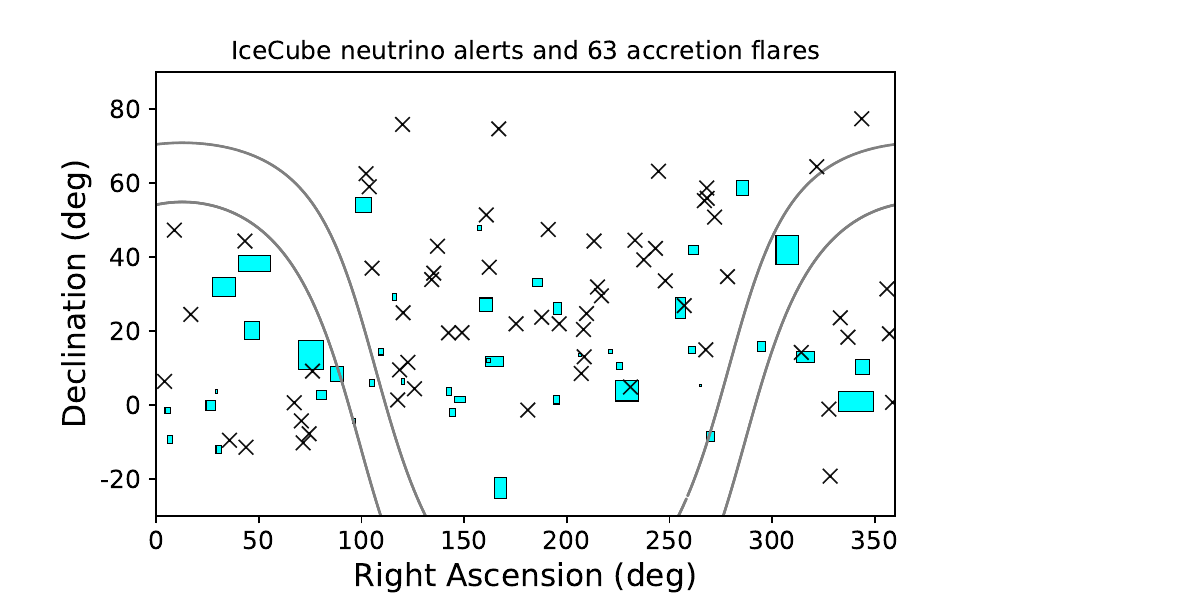}
\includegraphics[width=0.48\textwidth]{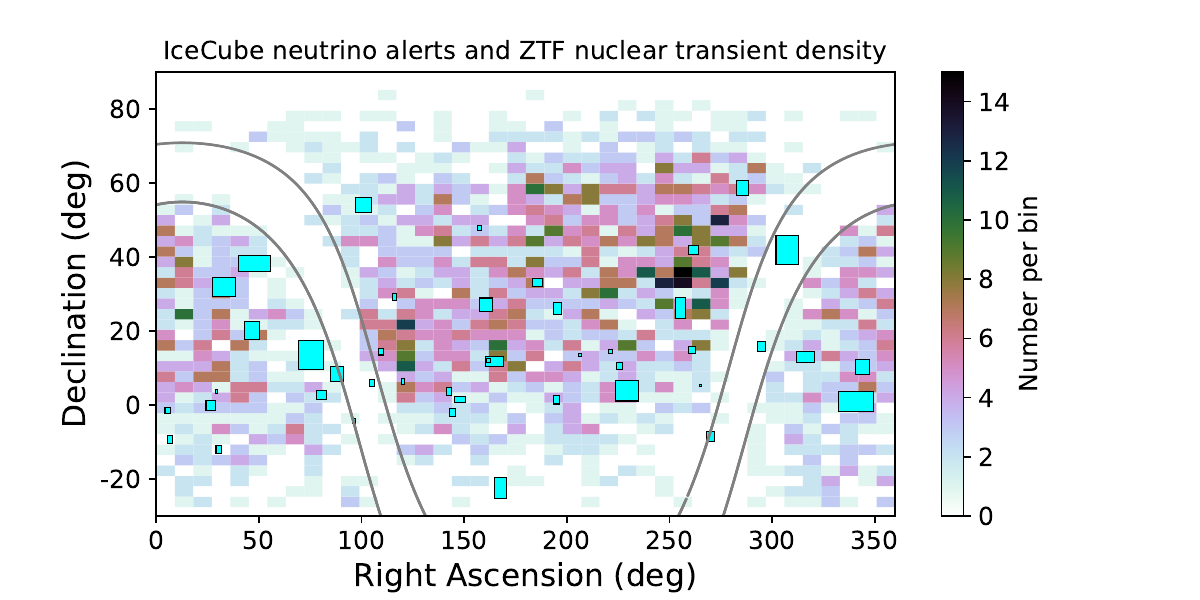}
\includegraphics[width=0.48\textwidth]{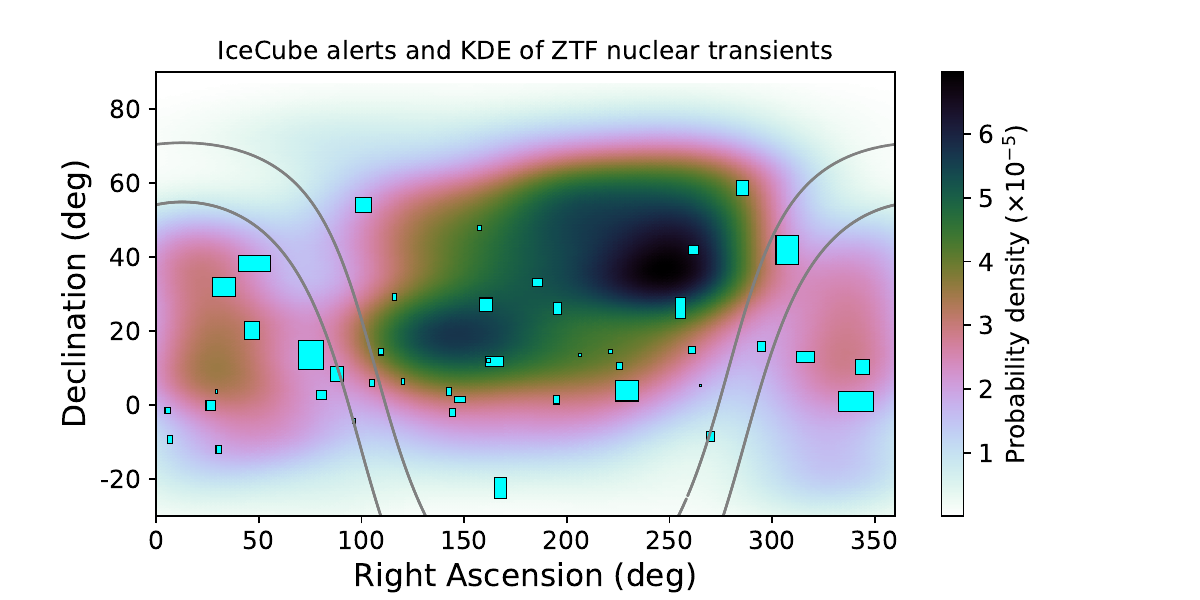}
\includegraphics[width=0.48\textwidth]{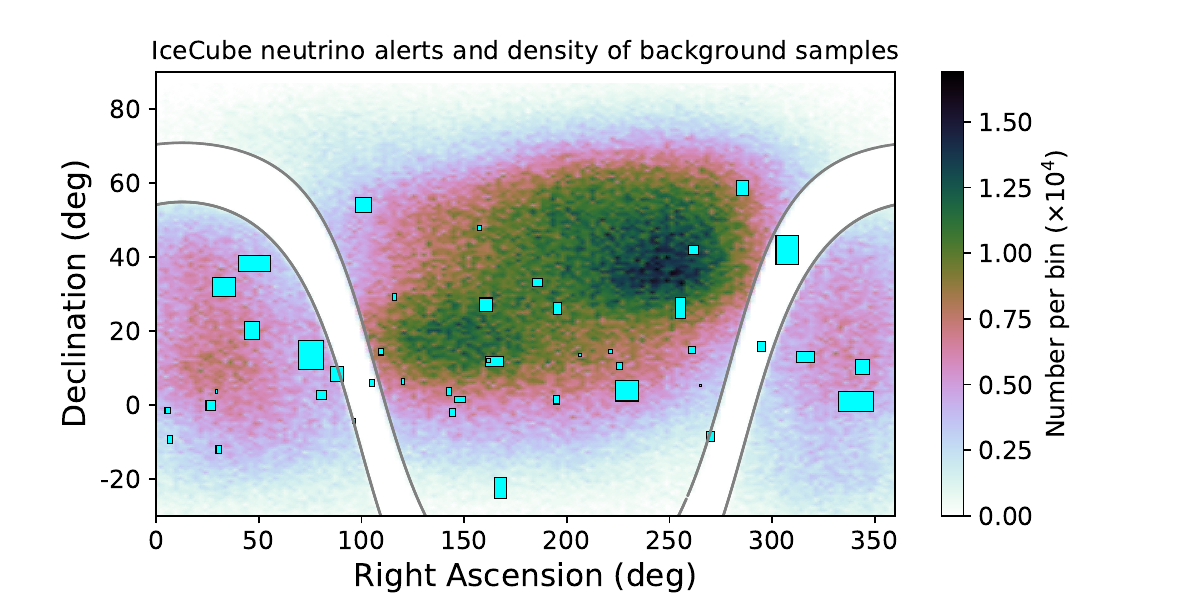} \\[25pt]
\includegraphics[width=0.29\textwidth]{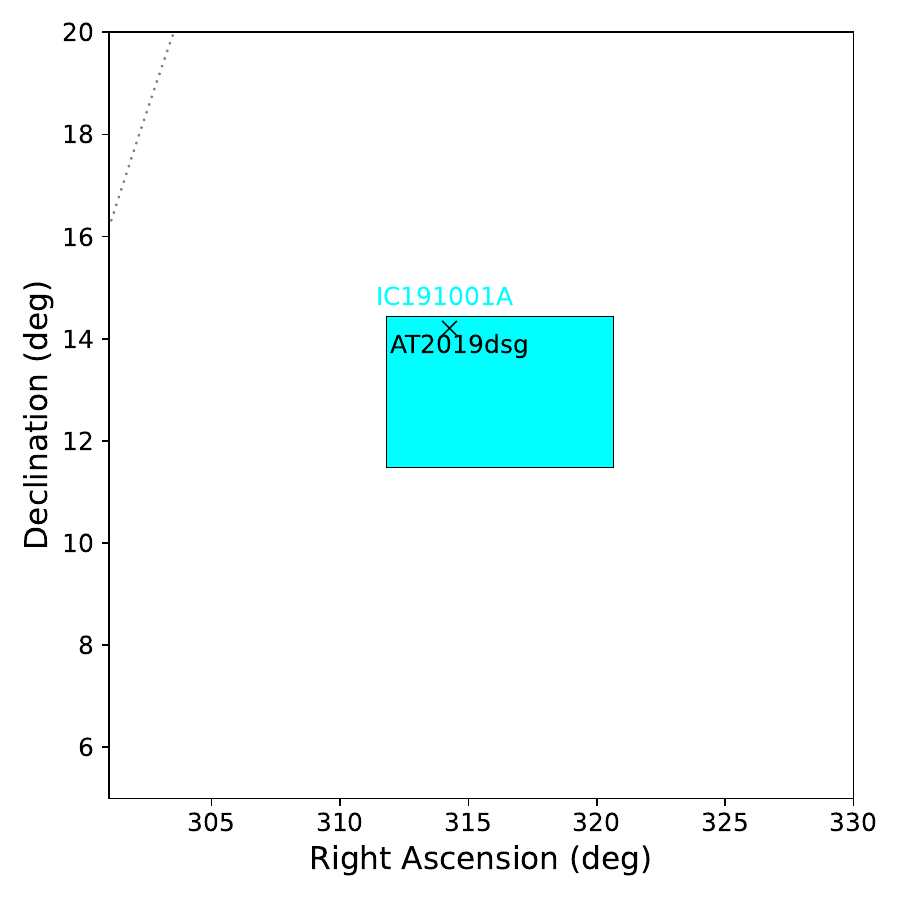}
 \includegraphics[width=0.29\textwidth]{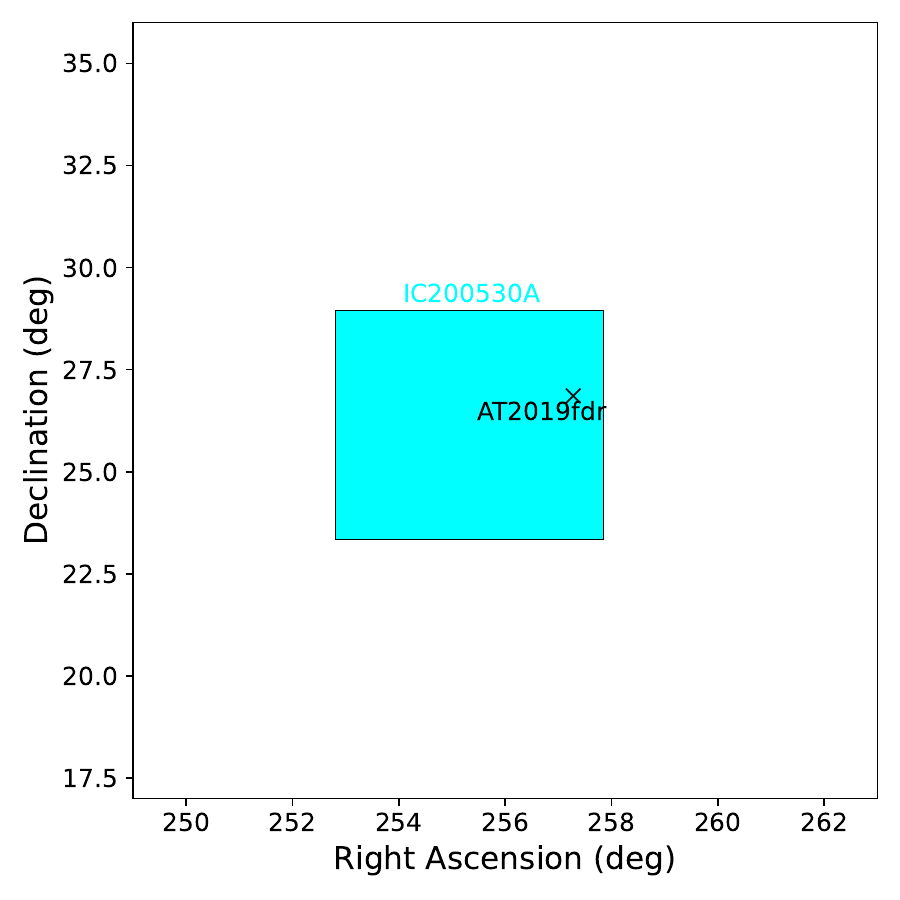}
 \includegraphics[width=0.29\textwidth]{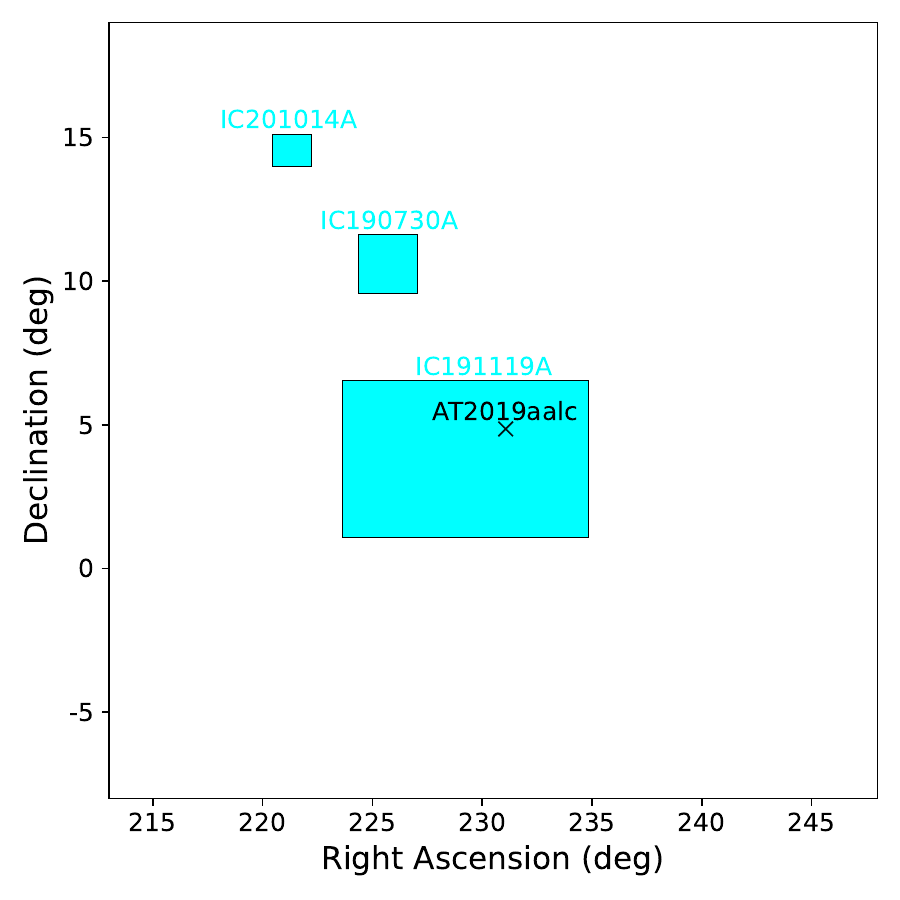}
  \caption{Sky maps of ZTF nuclear transients and IceCube neutrino alerts. In each panel, the cyan boxes indicate the IceCube 90\%CL reconstruction areas. The top-left map shows the 63 accretion flares with potential dust echoes that are used in this study. The top-right map shows the density distribution of our ``parent sample" of 3142 nuclear flares. The middle-left map shows the result of applying a Gaussian kernel density estimation (KDE) to this sample. The middle-right map shows the resulting sky distribution of Monte Carlo samples that are used for the background hypothesis in our likelihood method. These samples are drawn from the KDE, with the additional exclusion of the Galactic plane ($|b|>8$, as shown by the grey line). A linear color scale is used in all these panels. The three figures on the bottom row show close ups of the location of the three accretion flares coincident with an IceCube alert. }\label{fig:skymaps}
\end{figure*}

To conclude, the signal likelihood of a single neutrino and accretion flare coincidence is given by: 
\begin{equation}
    P(x | S) = 0.9
    \times  P(\Delta F_{\rm IR}/F_{\rm rms}|S) \times P(\Delta F_{\rm IR}  | S) 
\end{equation}

\subsubsection{Components of the background hypothesis}
We now consider the probability of observing the data under the null hypothesis. We again have two distinct factors:
\begin{enumerate}
\label{item:bg}
    \item The probability of a spatial and temporal coincidence if the neutrino was not produced by any accretion flare: $P_{\rm coin}(\alpha,\delta, t|B)$.

    \item The probability that the properties of an  accretion flare found in coincidence would be observed if the neutrino was not produced by that accretion flare. 
\end{enumerate}

The first item in this list accounts for the so-called chance coincidences, both spatially and temporally. The expectation value for the number of chance coincidences within the 90\%CL area ($\Omega_\nu$) of a given neutrino is given by $n_{\rm bg} \times \Omega_\nu$, with $n_{\rm bg}$ being the areal density of coincident events that are  expected for the background hypothesis. Because this expectation value for a single neutrino is always $\ll1$, the Poisson probability of obtaining a single spatial and temporal chance coincidence between a neutrino and an accretion flare can be written as 
\begin{eqnarray}
P_{\rm coin}(\alpha,\delta, t|B)=n_{\rm bg} ~ \Omega_\nu
\end{eqnarray}

A Monte Carlo approach is used to estimate $n_{\rm bg}$. We need to assign a peak time and location for each simulated accretion flare. Since our window for temporal coincidence is broad, we can use a non-parametric method to assign a time of peak, namely shuffling the peak times for each Monte Carlo realization. Next, we need a method to simulate celestial coordinates of ZTF extra-galactic transients. We start from the parent sample of 3142 nuclear transients with at least two post-peak NEOWISE observations (see section~\ref{sec:ztf}). The areal density in this sample is too low to allow a non-parametric approach. We therefore applied Gaussian KDE to obtain a continuous two-dimensional PDF of the celestial coordinates. To ensure the lack of events from the Galactic Plane is properly reflected, we enforce zero probability for Galactic coordinates with $|b|<8$~deg. After this small modification, we can use the resulting PDF to simulate celestial coordinates of ZTF nuclear transients using rejection sampling. These steps are summarized in Fig.~\ref{fig:skymaps}. After simulating $10^6$ samples of 63 accretion flares, we find a total of $4.7\times 10^5$ coincident neutrinos. Hence the total number of expected coincident events in a sample of 63 accretion flares is 0.48. To obtain the areal density of background coincidences, we divide this expectation value by $\Sigma \Omega_\nu=698.6$~deg$^2$, the total neutrino area that overlaps the ZTF footprint: $n_{\rm bg} = 0.48 / \Sigma \Omega_\nu=6.9\times 10^{-4}$~deg$^{-2}$.

The second probability in the list above can be found by considering the properties of accretion flares expected for a chance coincidence. Such accretion flares will be drawn at random from the general population. The probability to detect a given dust echo flux can be estimated from the flux distribution of the parent population of all nuclear transients with detected transient infrared emission. The probability to detect a given echo strength follows from the distribution of nuclear transients with a potential echo. Because the TDE candidate population dominates the upper-end of the echo strength distribution, we exclude these flares when applying a Gaussian KDE to obtain $P(\Delta F_{\rm IR}/F_{\rm rms}|B)$. The resulting PDFs are displayed in Figs.~\ref{fig:h_strength} \& \ref{fig:h_flux}.

%Such accretion flares will be drawn at random from the general population, with no preference towards higher average flux or higher echo strength. 
%We obtain this PDF using a KDE applied to the nuclear transients in the parent sample that are not classified as TDE candidates (see Figs.~\ref{fig:h_strength} and \ref{fig:h_flux}). 

\subsubsection{Final result: weighted likelihood ratio test}
In principle, the likelihood ratio test could be extended to include the probability of observing the neutrino properties detected at the IceCube Neutrino observatory for the background and signal hypothesis.
%, if the neutrinos was produced by the accretion flares or under the null hypothesis. 
This would account for the fact that some detector properties, such as detected energy, can be used to infer the probability of a neutrino being astrophysical. However, this information is not provided by IceCube. Instead, IceCube provides an estimated astrophysical probability for each neutrino event based on assumed properties of the astrophysical neutrino flux ($P_{\rm astro}$, see section~\ref{sec:IC}). We therefore choose to simply use this astrophysical probability as a weight in our likelihood ratio test.

We now collect the signal and background terms to define our test statistic: 
\begin{eqnarray}
    {\rm TS}\, &=& -2\ln \Bigg(
    \frac{ \hat{\mathcal{L}_{s}}} {\mathcal{L}_{0}}\Bigg) \nonumber \\
            &=&-2 \ln \Bigg( \prod_i P_{\rm astro} \times \frac{0.9}{\Omega_\nu n_{\rm bg}} \times  \nonumber \\
              &~&\frac{P(\Delta F_{\rm IR}/F_{\rm rms}  | S)}{P(\Delta F_{\rm IR}/F_{\rm rms}  | B)} \times   \frac{P(\Delta F_{\rm IR}  | S)}{P(\Delta F_{\rm IR}|B)} \Bigg) \label{eq:TS}
\end{eqnarray}
The sum runs over all neutrinos and the flare properties are evaluated for the (best-fit) accretion flare that is found in spatial and temporal coincidence with the neutrino.

\begin{figure}
  \centering
  \includegraphics[width=0.5\textwidth]{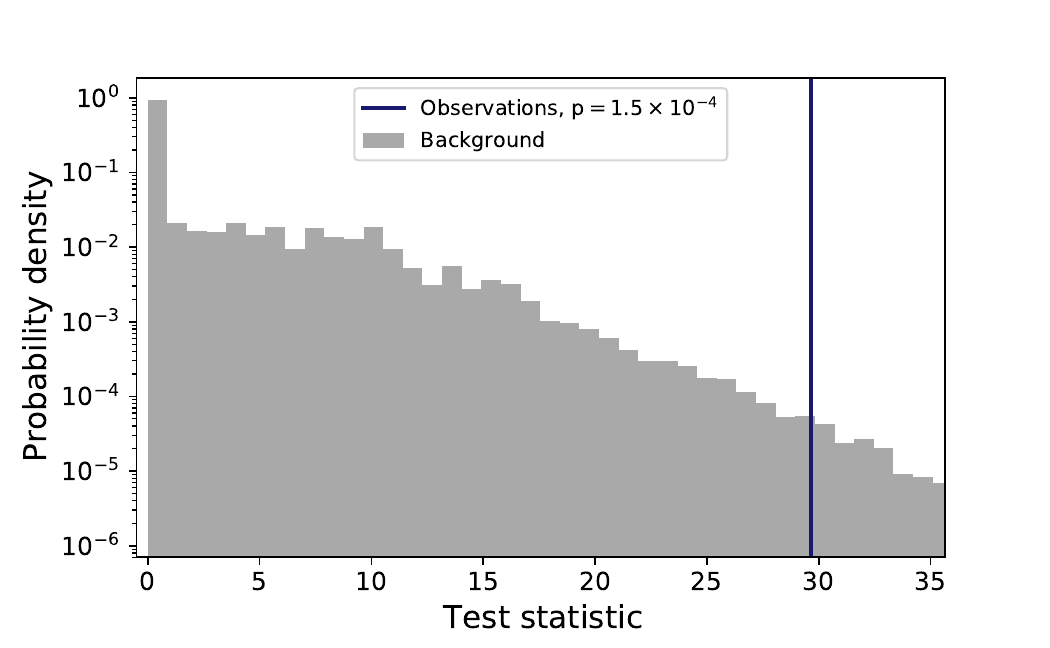}
  \caption{The distribution of the test statistics after redistributing the accretion flares in the ZTF sky. From this Monte Carlo simulation we find a probability $p=1.5\times10^{-4}$ that the three accretion flares we associate with a high-energy neutrino are explained by a chance coincidence. The peak at zero is due to Monte Carlo realizations that yield zero coincident events. }
  \label{fig:TS}
\end{figure}

For the observed sample of 63 accretion flares we find TS$=29.6$. 
We can now estimate the significance of this result by computing the TS distribution under the background hypothesis. We use 10$^6$ samples of 63 accretion flares drawn from the PDF of the celestial coordinates of nuclear transients (see Fig.~\ref{fig:skymaps}), with shuffled flare start times. For each of these Monte Carlo realizations, we compute the TS using the coincident neutrino and flare pairs. The resulting distribution of TS is shown in Fig.~\ref{fig:TS}. A fraction $1.5 \times 10^{-4}$ of the simulations for the background hypothesis have an equal or greater TS than the observations, corresponding to a significance of $3.6\sigma$. 

The detection of a third neutrino in coincidence with an accretion flare (i.e., \lancel), decreases the probability of the background hypothesis by a factor 60; if our search had not uncovered this new event, the significance would have been 2.4$\sigma$. 

\subsection{Statistical power}\label{sec:power}
It can be instructive to investigate the statistical power of our likelihood method (Eq.~\ref{eq:TS}) as a function of the expectation value of the number of signal neutrinos, $n_{\rm exp}$. To simulate a signal for a given $n_{\rm exp}$, we draw $M$ events from a Poisson distribution with expectation value $n_{\rm exp}$. Each of these $M$ events has a 90\% probability to be detected in coincidence with an accretion flare. To each simulated neutrino-flare pair we assign an echo flux and strength from the PDF of the signal hypothesis (see Figs.~ \ref{fig:h_strength} \& \ref{fig:h_flux}). We repeat this signal simulation 10,000 times to obtain the TS distribution for a given $n_{\rm exp}$. We compare this distribution to a few critical TS values: ${\rm TS}=0$, which is the median of the background TS distribution, and ${\rm TS}=27.9$, ${\rm TS}=43.8$, and ${\rm TS}=57.5$ which correspond to the threshold for a 3, 4, and 5$\sigma$ significance, respectively (the limiting TS value for $5\sigma$ is obtained by extrapolating the background TS distributing, using the approach outlined in \citealt{Aartsen17}). 

\begin{figure}
  \centering
\includegraphics[width=0.5\textwidth]{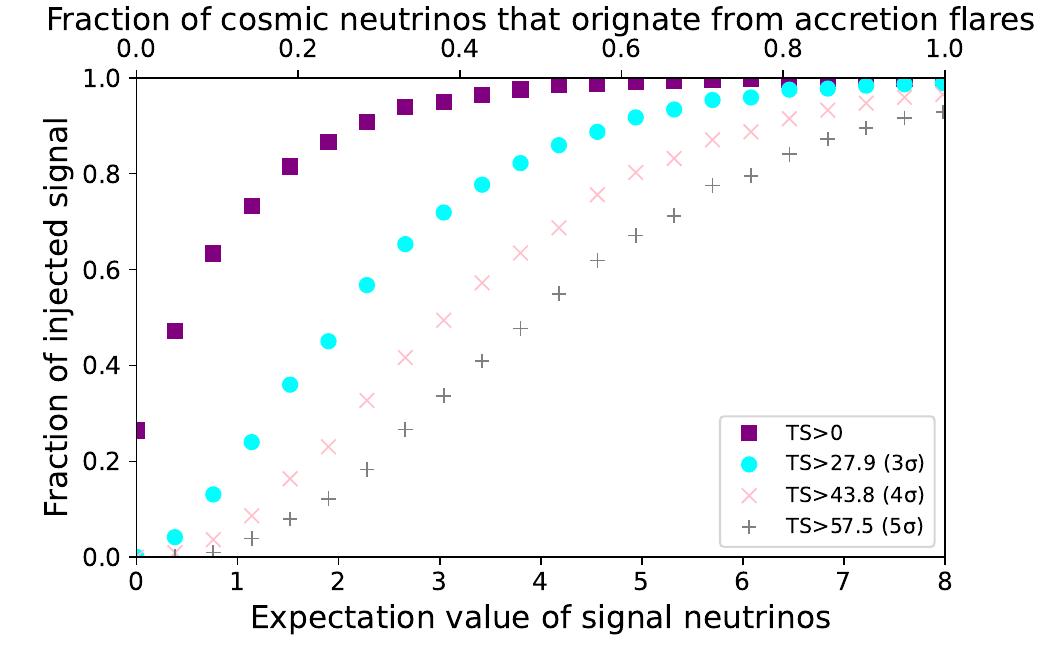}
\caption{Assessment of the statistical power of our weighted likelihood ratio test (Eq.~\ref{eq:TS}). The horizontal axis encodes the expectation value for the number of neutrinos that are detected in coincidence with an accretion flare in our sample ($n_{\rm exp}$ in Eq.~\ref{eq:fcosmic}). For each expectation value, one can simulate a distribution of the test statistic (TS; Eq.~\ref{eq:TS}) and compute the fraction of injected signal trials above a given TS threshold value. The top axis shows the fraction of astrophysical IceCube alerts that originate from accretion flares in our catalog (based on Eq.\ref{eq:fcosmic}). For reference, based on the detection of three coincident neutrino-flare pairs, the 90\%CL range of $n_{\rm exp}$ is (1.1, 6.8).}\label{fig:statpower}
\end{figure}

When at least 50\% of the simulated signal trials have ${\rm TS}>0$, our test statistic can be called admissible (on average, we have enough sensitivity to prefer the signal hypothesis over the null hypothesis); this happens when $n_{\rm exp}>0.3$. For reference, our detection of three coincident neutrinos implies $n_{\rm exp}=3_{-1.9}^{+3.8}$ (90\% CL). From Fig.~\ref{fig:statpower} we see that for this range of $n_{\rm exp}$, a significance between 3 and 4$\sigma$ should be expected.

The number of signal neutrinos is related to the fraction of cosmic high-energy neutrinos that originate from accretion flares, $f_{\rm flare}$:

\begin{align}
    f_{\rm ac} = \frac{n_{\rm exp}} {N_{\rm cosmic}\eta_{\rm ZTF}} \label{eq:fcosmic}
\end{align}

Here $N_{\rm cosmic}\approx 16$ is the number of astrophysical neutrino alerts that are included in our search (see Section~\ref{sec:IC}). The parameter $\eta_{\rm ZTF}$ accounts for the fraction of astrophysical neutrinos from accretion flares that detected by IceCube, but not included in our catalog. In particular, a large population of relatively high-redshift accretion flares ($z > 0.5$) will 
not be detected in ZTF or NEOWISE, but these can yield a sizeable fraction of neutrino alerts. Following \citet{Stein20}, we adopt $\eta_{\rm ZTF}=0.5$. The resulting $f_{\rm ac}$ is shown at the top of 
Fig.~\ref{fig:statpower}.

\subsection{Cross-checks and robustness}\label{sec:robust}
\begin{figure}
  \centering
  \includegraphics[width=0.5\textwidth]{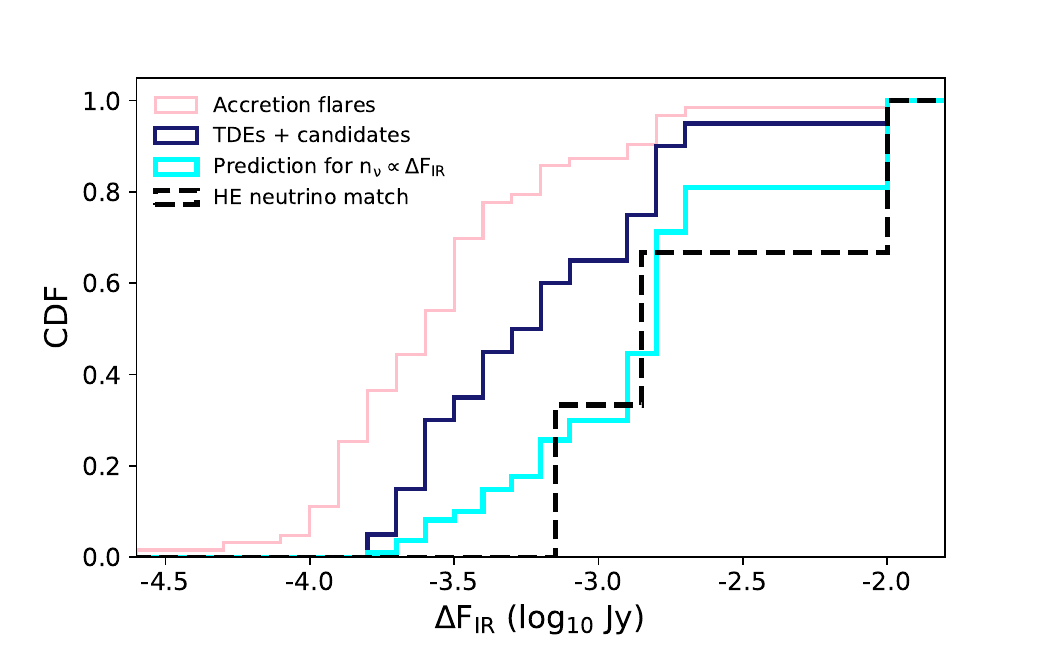}
  \caption{Cumulative distribution of the dust echo flux. Here we check the assumption that $P(\Delta F_{\rm IR}|S)\propto \Delta F_{\rm IR}$ for the signal hypothesis of the likelihood method (Eq.~\ref{eq:TS}). For each of the 18 TDE candidates (i.e., accretion flares with $M_{\rm BH}<10^8\,M_\odot$) we predict the neutrino expectation value using a linear coupling with echo flux (normalized to yield 1 neutrino for a echo flux of 0.1~mJy). By drawing from a Poisson distribution we simulate neutrino detections from this population. The predicted echo flux distribution for samples that yield three neutrino detections from three different flares (cyan curve) is consistent with the observed flux of the three neutrino-detected flares (black dashed line).  }
  \label{fig:h_flux_predict}
\end{figure}

Now that we have established evidence for three accretion flares with a neutrino counterpart, we can do a cross-check on the assumption that the neutrino flux is coupled linearly to the infrared echo flux. We first make an ansatz for the coupling strength between the neutrino expectation value ($n_\nu$) and the infrared flux: $n_\nu = \epsilon \Delta F_{\rm IR}$, with $\epsilon=100 {\rm Jy}^{-1}$. Applying this coupling to the observed echo flux distribution of the TDE candidates we obtain $n_\nu$ for each TDE candidate. After simulating neutrino detections from each TDE by drawing from a Poisson distribution, we find that, for this coupling strength, 36\% of the simulations yield  at least three difference sources, each with at least one neutrino. In most cases, we obtain one neutrino per source (e.g., for the subset of simulations with exactly three neutrino sources, 46\% yield more than one neutrino from a single flare). In Fig.~\ref{fig:h_flux_predict} we show the distribution of echo flux for the simulations that yield exactly three detected neutrinos from three different sources. This prediction is consistent with the flux distribution of the three flares that have evidence for a neutrino counterpart.

While this agreement is encouraging, we should expect deviations from some of the simplifying assumptions that are made to construct the test statistic (Eq.~\ref{eq:TS}). This will decrease the sensitivity of our statistical method. Below we test the robustness of our significance estimate by making several alternative choices for the input of Eq.~\ref{eq:TS}. 

First, the simulation of the expected echo flux distribution for a linear coupling with the neutrino expectation value and infrared flux could also be considered as input for $P(\Delta F_{\rm IR}|S)$. This approach is not ideal for two reasons: (1) the coupling strength is a free parameter; (2) the numbered of observed coincident events is needed to normalize this PDF. Nevertheless, if we redo our estimate of the significance using the predicted dust echo flux distribution as shown in Fig.~\ref{fig:h_flux_predict}, we obtain $p=1.7\times 10^{-4}$ for the background hypothesis (3.6$\sigma$).

Next, we consider the  PDF of the celestial coordinates of nuclear transients, which is required to generate the distribution of background flares in our Monte Carlo analysis. The resulting significance is not sensitive to the details of this PDF. To demonstrate this, we can draw the background Monte Carlo samples from the observed local galaxy distribution, without KDE smoothing. We use the 2MASS redshift survey \citep{Huchra12} and select all galaxies that fall within the declination limits of ZTF ($-28<\delta<86$), yielding $3.1\times 10^4$ sources. The resulting sky distribution is uniform within the ZTF footprint, which is not consistent with the observed distribution of nuclear transients (Fig.~\ref{fig:skymaps}). While the 2MASS galaxy distribution is clearly not an appropriate description of the celestial coordinates of our nuclear flare sample, this has a modest influence on the inferred significance if this map is selected to generate background samples. After drawing the coordinates of the background samples from the 2MASS galaxies, we find $n_{\rm bg}=8.8\times 10^{-4}$ and $p=2.1\times 10^{-4}$ (3.5$\sigma$).  

Finally, we consider the impact of the selection of accretion flares based on the ZTF properties of the light curve. If we make no cuts on the ZTF properties and simply select all potential dust echoes (i.e., $\Delta F_{\rm IR}/F_{\rm rms}>F_{\rm rms}/\sigma_F$) we obtain 163 sources.  For this sample, we expect 1.9 matches under the background hypothesis (compared to 0.48 for the original sample of 63 accretion flares). This larger population yields one additional coincident neutrino (for this fourth source, ZTF19aaozcxx, the background hypothesis is preferred over the signal hypothesis, TS$_i=0$). The probability of the null hypothesis is $p=3.2\times 10^{-4}$ (3.4$\sigma$).

\subsection{Density-based estimate of significance}\label{sec:dens}
As a final cross-check on the robustness of our likelihood and Monte Carlo method we consider a simple, yet instructive estimate of the significance based solely on the areal density of accretion flares with large dust echoes. For this estimate, no Monte Carlo sampling is needed. We first note that for $\Delta F_{\rm IR}/F_{\rm rms}>10$, the flare population is dominated by TDE-like echoes (Fig.~\ref{fig:df_mbh}). Applying this cut on echo strength leaves 29 flares. The fraction of neutrino alerts that yield a temporal coincidence with these flares is $f_{\rm temp}=0.24$. We thus obtain the effective source density of large echoes: $n_{\rm eff}=29 \times f_{\rm temp}/\Omega_{\rm ZTF}=2.5\times 10^{-4}$~deg$^{-2}$, with $\Omega_{\rm ZTF}=2.8\times 10^4$~deg$^2$ the extragalactic sky seen by ZTF \citep{Stein20}. Multiplying this effective source density with the total area of the IceCube neutrino alerts,
%($\Sigma \Omega_\nu=698.6$~deg$^2$)
we obtain the expected number of chance coincidences. This expectation value is $0.17$ and the Poisson probability to see at least three events is $8\times 10^{-4}$ (3.2$\sigma$).

\section{Multi-wavelength properties}\label{sec:mm}

\begin{table*}
\centering
    \begin{tabular}{lcccccccccccc}
        \hline
        Flare & Neutrino  & $z$  & $\Delta t$  & $\Delta d$ & $L_{\rm neutrino}$ & $T_X$  & $L_X$ & $L_{\rm radio}$ & $L_{\gamma}$ &  $L_{\rm bol}$ & $M_{\rm BH}$ & $f_{\rm Edd}$ \\ 
        & & & (days) & (deg) & (${\rm erg}\,{\rm s}^{-1}$) & (keV) & (${\rm erg}\,{\rm s}^{-1}$) & (${\rm erg}\,{\rm s}^{-1}$) & $ ({\rm erg}\,{\rm s}^{-1}$)  &  (${\rm erg}\,{\rm s}^{-1}$)& ($M_\odot$) & \\
        \hline\hline
        AT2019dsg&IC191001A&0.051&150&1.3&$<10^{44.5}$&$0.07\pm0.01$&$10^{43.4}$&$10^{38.9}$&$<10^{43.1}$&$10^{45.3}$&$10^{6.6}$&4.3\\
AT2019fdr&IC200530A&0.267&289&1.7&$<10^{45.8}$&$0.06\pm0.03$&$10^{43.1}$&$10^{39.3}$&$<10^{44.3}$&$10^{44.9}$&$10^{7.1}$&0.5\\
AT2019aalc&IC191119A&0.036&148&1.9&$<10^{44.2}$&$0.17\pm0.01$&$10^{42.1}$&$10^{38.7}$&$<10^{43.3}$&$10^{45.1}$&$10^{7.2}$&0.6\\

    \end{tabular}
    \caption{Multi-messenger inference. The time difference of the neutrino arrival ($\Delta t$) is measured relative to the optical peak of the light curve. The angular offset ($\Delta d$) is measured relative to the best-fit neutrino position (Table~\ref{tab:IC}). The upper limit on the neutrino luminosity is estimated by assuming an expectation value of one particle in $\Delta t$ (the true neutrino luminosity could be one or two orders of magnitude lower due to the Eddington bias, see Sec.~\ref{sec:energetics}).  The X-ray temperature ($T_X$) and luminosity ($L_X$) are based on SRG/eROSITA or {\it Swift}/XRT and cover an energy range of 0.2 to 10~keV (see Sec.~\ref{sec:SRG}). The radio luminosity is measured at $\nu \approx 5$~GHz (see Sec.~\ref{sec:radio}). The 95\% CL upper limit on the $\gamma$-ray luminosity ($L_\gamma$) is obtained from data of the {\it Fermi}-LAT telescope and covers photons in the energy range 0.1--800~GeV (see Sec.~\ref{sec:fermi}). The black hole mass ($M_{\rm BH}$) is estimated from the optical spectrum of the host galaxy (Sec.~\ref{sec:mbh}). The Eddington ratio ($f_{\rm Edd}$) follows from the bolometric luminosity ($L_{\rm bol}$) as estimated from the duration of the dust echo (Sec.\ref{sec:neoWISE}). }\label{tab:mm}
\end{table*}

\subsection{Detection at radio wavelengths}\label{sec:radio}
The first neutrino-detected source \bran, showed transient radio emission, evolving on a timescale of month, with a peak luminosity ($\nu L_\nu$) of $8 \times 10^{38}\,{\rm erg}\,{\rm s}^{-1}$ at 10~GHz \citep{Stein20}. The source \tywin was detected in radio follow-up observations with similar luminosity ($2 \times 10^{39}\,{\rm erg}\,{\rm s}^{-1}$ at 10 GHz), with marginal evidence for variability at 10~GHz frequencies \citep{Reusch22}. Due to its higher redshift, the radio emission from \tywin is relatively faint (0.1~mJy) and below the detection threshold of wide-field radio surveys such as FIRST \citep{Becker95} or VLASS \citep{Lacy20}. 

Finally, the third accretion flare coincident with a high-energy neutrino, \lancel, is detected in both FIRST and VLASS. The FIRST observation were obtained in the year 2000, thus predating the optical flare by almost two decades. These radio observations yield a luminosity of $2\times 10^{38}\,{\rm erg}\,{\rm s}^{-1}$ at 1.4~GHz. The VLASS observations were obtained on two dates, 2019-03-14 and 2021-11-06, yielding 3~GHz radio luminosities of $3\times 10^{38}$ and $5\times 10^{38}$ \,${\rm erg}\,{\rm s}^{-1}$, respectively. This factor $\approx 2$ flux increase from the first VLASS epoch (three months before the optical peak) to the second VLASS epoch (2 years post-peak) is statistically significant (at the 8$\sigma$-level, as estimated from the rms in the VLASS ``Quick Look" images). 

\subsection{\textit{Fermi} gamma-ray upper limit}\label{sec:fermi}
We analysed data from the \textit{Fermi} Large Area Telescope (\textit{Fermi}-LAT; \citealt{2009ApJ...697.1071A}), a pair-conversion telescope sensitive to gamma rays with energies from 20 MeV to greater than 300 GeV.
Following the approach outlined in \citet{Stein20}, we use the photon event class from Pass 8 \textit{Fermi}-LAT data (P8R3\_SOURCE), and select a 15 $\times$ 15 deg$^2$ region centered at the target of interest, with photon energies from 100 MeV to 800 GeV. We use the corresponding LAT instrument response functions P8R3\_SOURCE\_V2 with the recommended spectral models \textit{gll\_iem\_v07.fits} and \textit{iso\_P8R3\_SOURCE\_V2\_v1.txt} for the Galactic diffuse and isotropic component respectively, as hosted by the \href{https://fermi.gsfc.nasa.gov/ssc/data/access/lat/BackgroundModels.html}{FSSC}.  We perform a likelihood analysis, binned spatially with 0.1 deg resolution and 10 logarithmically-spaced bins per energy decade, using the \textit{Fermi}-LAT ScienceTools package (fermitools v1.2.23) along with the \textit{fermipy} package v0.19.0 \citep{2017ICRC...35..824W}.

We studied the region of \lancel in the time interval that includes 207 days of observations from the discovery of the optical emission on 2019 April 26 to the observation of the high-energy neutrino IC191119A on 2019 November 19. 
In this time interval, there is no significant ($\geq 5 \sigma$) detection for any new gamma-ray source identified with a localization consistent with IC191119A. Two sources from the fourth \textit{Fermi}-LAT point source catalog (4FGL-DR2; \citealt{2020ApJS..247...33A,2020arXiv200511208B}), consistent with the IC191119A localization, are detected in this interval. These are 4FGL J1512.2+0202 (4.1 deg from IC191119A), associated with the object PKS 1509+022, and 4FGL J1505.0+0326 (5.0 deg from IC191119A), associated with the object PKS 1502+036. The flux values measured for these two detections are consistent with the average values observed in 4FGL-DR2.

Likewise, we studied \textit{Fermi}-LAT data of \tywin{} using a time window from its discovery on 2019 May 3 to the arrival time of IC200530A on 2020 May 30. In this window, no gamma-ray sources were significantly ($\geq 5 \sigma$) detected within the localization region of IC200530A, including both previously known 4FGL-DR2 catalogued sources and new gamma-ray excesses.

%no known 4FGL-DR2 source and no new gamma-ray excess are significantly ($\geq 5 \sigma$) detected within the localization region of IC200530A. % 

For both \tywin and \lancel, we test a point-source hypothesis at their position under the assumption of a power-law spectrum. The 95\% CL upper limit for the energy flux (i.e., integrated over the whole analysis energy range) listed in Table~\ref{tab:mm} is derived for a power-law spectrum ($dN/dE \propto E^{-\Gamma}$) with photon power-law index $\Gamma=2$. The {\it Fermi}-LAT upper limit for \bran listed in Table~\ref{tab:mm} is obtained using the same LAT data analysis strategy and covers a similar time window ($150$~days) relative to the optical discovery and the neutrino arrival time \citep{Stein20}. 

No significant gamma-ray emission is detected in \textit{Fermi}-LAT data at the source positions prior to their optical discoveries \citep{2017ApJ...846...34A}.

\subsection{SRG/eROSITA X-ray detections}\label{sec:SRG}
The SRG X-ray observatory \citep{sunyaev21} was launched to the halo orbit around Sun-Earth L2 point on 2019 July 13. On 2019 December 8, it started the all-sky survey, which will eventually comprise eight independent 6-month long scans of the entire sky. In the course of the sky survey, each point on the sky is visited with a typical cadence of 6 months (the exposure and number of scans depends on ecliptic latitude). The eROSITA soft X-ray telescope \citep{predehl20} operates in the 0.2-9 keV energy band with its effective area peaking at $\approx 1.5$ keV.% 
%where it  achieves $\sim 950$ cm$^2$ (survey). 

As of October 2021, \lancel (= SRGe J152416.7+045118) has been visited four times starting from February 2,  2020 with 6 month intervals and was detected in each scan. The X-ray light curve of the source as seen by eROSITA reached a plateau between 2020 August and 2021 January with the $0.3-2.0$ keV flux of $\approx 4.6\times 10^{-13}$~erg\,s$^{-1}$cm$^{-2}$. The source had a soft thermal spectrum with the best fit blackbody temperature of $kT=172\pm 10$ eV.

The source \tywin (= SRGe J170906.6+265124) has been visited four times starting 2020 from March 13. The sources was detected only once, on 2021 March 10--11, with a 0.3--2.0 keV  flux of $\approx 6.0\times 10^{-14}$~erg\,s$^{-1}$cm$^{-2}$ and an extremely soft thermal spectrum with a temperature of $56_{-26}^{+32}$ eV. This flare displayed the softest X-ray spectrum of all TDEs detected by eROSITA so far \citep{sazonov21}. 
In the three visits when the source remained undetected, the upper limit on its flux  was in the range of  $\sim (2-5)\times 10^{-14}$~erg\,s$^{-1}$cm$^{-2}$, see \citet{Reusch22} for details. 

The source \bran has been visited by eROSITA three times starting from 2020 May 9 and so far remained undetected with the $3\sigma$ upper limit for the combined data of the three visits of $\lesssim 1.9\times 10^{-14}$~erg\,s$^{-1}$cm$^{-2}$ (assuming a power law spectrum with the slope of $\Gamma=1.8$).
%The forth visit of the source is expected to take place in mid November 2021. 
The X-ray measurements of \bran listed in Table~\ref{tab:mm} are based on {\it Swift}/XRT; \citealt{Gehrels04})  observations that were obtained closer to the optical peak of the flare \citep{vanVelzen20,Stein20} than the eROSITA observations.

\section{Results and implications}\label{sec:impl}

\subsection{A new population of neutrino sources}
By using infrared observations of dust echoes as a tracer of large accretion events near black holes, we are able to unify TDEs in quiescent galaxies and (extreme) AGN flares. This allows us to test a new hypothesis: large amplitude accretion flares are sources of high-energy neutrino emission. Thanks to our systematic selection of dust echoes, we can use a large sample of 40 neutrinos, compared to the 24 neutrinos that were followed-up by ZTF to find \bran and \tywin \citep{Stein20, Reusch22}. This larger sample allows us to uncover a third flare (ZTF19aaejtoy/\lancel), which happens to have the highest dust echo flux of all ZTF transients. The significance of this population of three flares is 3.6$\sigma$. The detection of the third flare reduces the probability of a chance association by a factor of $\approx60$.

Additional evidence for neutrino emission from accretion flares follows from the shared multi-wavelength properties of the three sources with a neutrino counterpart.  

For \lancel, we measure a soft thermal spectrum with temperature of $kT = 172 \pm 10$~eV.  Such soft thermal emission is rare: of all accretion flares with potential dust echoes, less than $13$\% are as soft as \lancel ($kT < 172$~eV), and even fewer are as soft as \bran and \tywin. These distinctive X-ray properties provide 3$\sigma$-level ($p = 0.13^3$) evidence for the hypothesis that accretion flares are correlated with high-energy neutrinos.

Another shared property of the three events is low-luminosity radio emission (see section~\ref{sec:radio}).
In our sample of accretion flares with dust echoes, less than 10\% are detected in archival radio observations (FIRST or VLASS) and a similarly low fraction of TDEs is detected in radio follow-up observations \citep{Alexander20}. If the three neutrino associations in our sample of accretion flares are due to chance, the probability to find three radio detections is $<10^{-3}$.    

%The pre-flare radio detections for \lancel imply that, similar to \tywin \citep{Reusch22}, the radio emission is not directly related to the optical flare or the neutrino. Nevertheless, this optically thin radio emission implies that the black hole was able to produce an outflow in its recent past and could therefore indicate an accumulation of substantial magnetic flux near the black hole horizon \citep{Tchekhovskoy11}. Not all accreting black holes are able to reach this accretion state \citep{Tchekhovskoy11}.

\begin{figure*}
  \centering
  \includegraphics[trim=0mm 13mm 0mm 8mm, clip, width=0.7\textwidth]{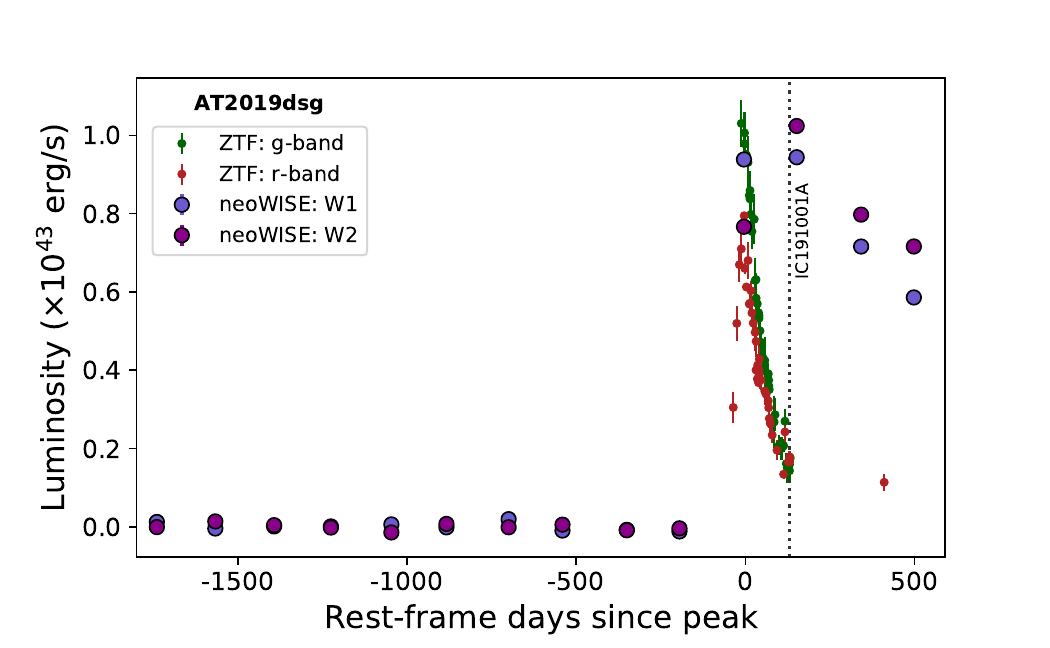}\\
  \includegraphics[trim=0mm 13mm 0mm 8mm, clip, width=0.7\textwidth]{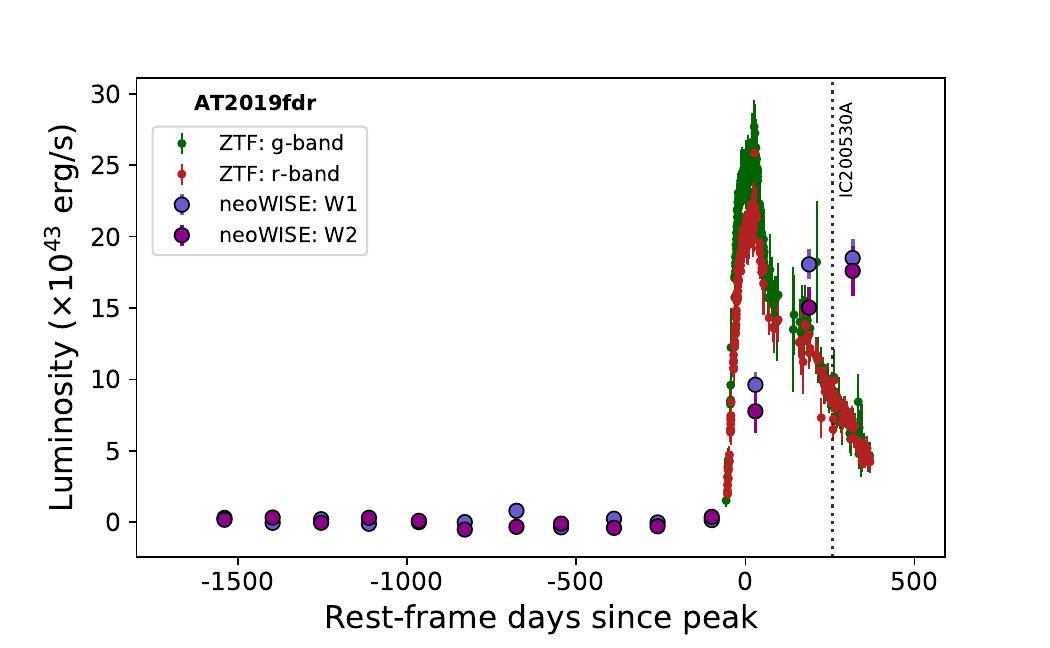}\\
  \includegraphics[trim=0mm 0mm 0mm 8mm, clip, width=0.7\textwidth]{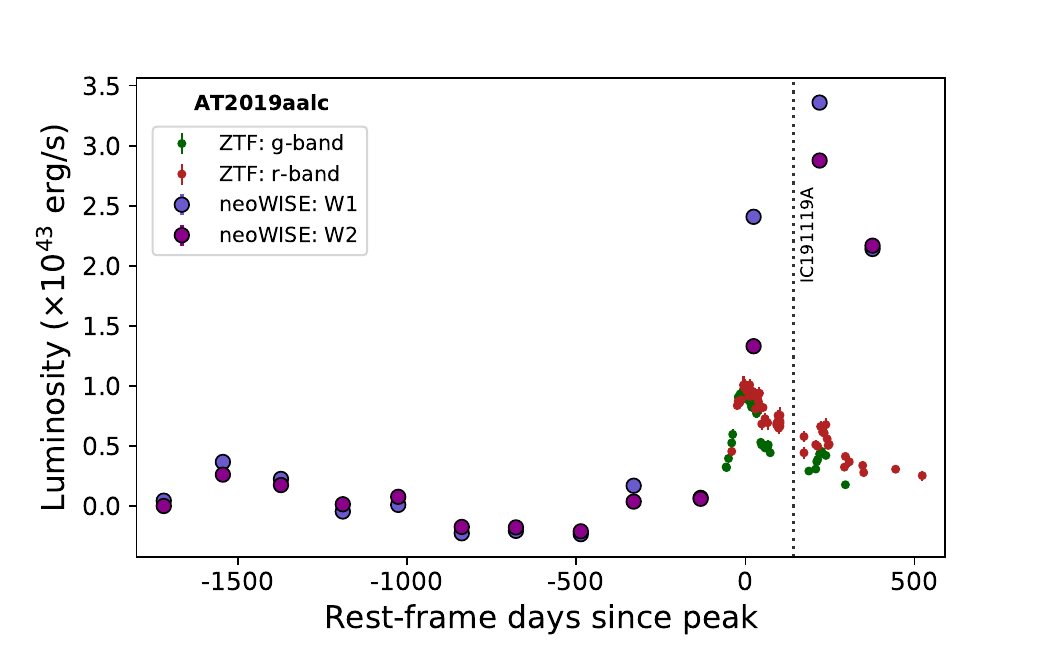}
  \caption{{Neutrino detections for three accretion flares.} For each source, the neutrino arrived (dotted vertical lines) a few months after the peak of the optical light curve (red and green symbols for the ZTF $g$ and $r$-band, respectively). The infrared light curve (blue and purple symbols for the WISE W1 and W2, respectively) evolves on longer timescales due to the large distance of the dust sublimation radius ($\sim 0.1$~pc). From the duration of the dust reverberation signal we infer a peak luminosity near the Eddington limit for all three flares (Table~\ref{tab:mm}). }
  \label{fig:lcs}
\end{figure*}

Transient radio emission and soft X-ray emission is not consistent with a supernova explanation for \tywin that has been proposed by \citet{pitik_22}.
Taken together, the shared X-ray, radio, and optical properties of the three flares with neutrino coincidences (Fig.~\ref{fig:lcs}) point to a new population of cosmic particle accelerators powered by transient accretion onto massive black holes.   

\subsection{Energetics}\label{sec:energetics}
Around PeV energies, the expectation value for the number of  IceCube neutrino detections can be approximated as
\begin{equation}
    N_{\rm neutrino} = 0.7 \times \frac{E_{\rm neutrino}}{10^{51}~{\rm erg}} \left(\frac{d}{100~{\rm Mpc}} \right)^{-2}
    \label{eq:Nnu}
\end{equation}
with $E_{\rm neutrino}$ the total energy carried by (mono-energetic) neutrinos \citep{Stein20}. Our single-neutrino associations suffer from a significant Eddington bias \citep{Strotjohann16}. Since our analysis considered a sample of 63 accretion flares, we should expect $N_{\rm neutrino}<1$ for any individual source in our sample. This implies Eq.~\ref{eq:Nnu} cannot be inverted to estimate the energy in neutrinos emitted by a single source. Only a upper limit can be estimated by setting $N_{\rm neutrino}=1$, the result is shown in Table~\ref{tab:mm}. 

Instead of investigating the neutrino luminosity of a single source, a more useful approach is to demand that the expectation value for the number of neutrino associations for the full population of 63 accretion flares is equal to the observed value (i.e., three). To obtain $N_{\rm neutrino}$ for the entire population, we again assume a single coupling strength ($\eta_{\nu}$) between the electromagnetic energy and the energy carried by neutrinos:
\begin{equation}
    E_{\rm neutrino} = \eta_{\nu} E_{\rm bol} \quad.
    \label{eq:etanu}
\end{equation}
We estimate the bolometric energy of the flare from its bolometric luminosity as obtained from the duration of the dust echo ($L_{\rm bol}$, see Sec.~\ref{sec:neoWISE}) and the $e$-folding time $\tau$ of the decaying part of the optical light curve (see Fig.~\ref{fig:risefade}): $E_{\rm bol} = L_{\rm bol} \tau$. The total bolometric flare energy of the 63 accretion flares is $1\times 10^{54}$~erg. Hence an expectation of $N_{\rm neutrino} = 3_{-2}^{+4}$ (90\% CL) yields a mean coupling strength $\eta_\nu = 4_{-3}^{+5}\times 10^{-2}$.

For most models of particle acceleration in AGN accretion disks, the luminosity in high-energy cosmic rays is between one and two orders of magnitude larger than the neutrino luminosity \citep[][]{Begelman90,Murase20}. Our estimate of the fraction of the electromagnetic energy that is given to neutrinos ($\eta_\nu\sim 10^{-2}$) thus implies that the energy in cosmic rays could be of the same order as the bolometric luminosity emitted by the flare.

If we apply our simple model (i.e., a fixed coupling strength between the bolometric energy and the energy emitted in neutrinos) to individual sources, we find a neutrino expectation value of 0.08, 0.004, and 0.21 for \bran, \tywin, and \lancel, respectively. We note that for \bran, a model of particle acceleration in the core of an accretion disk with a constant accretion rate yields an expectation of 0.1 IceCube high-energy neutrinos for an integration time of 1 year \citep{Murase20}. We refer to \citealt{Winter23} for further exploration of potential particle acceleration mechanisms in all three accretion flares.

%\subsection{Origin of extreme flares from black holes}

\subsection{Neutrino arrival delay}
If the optical luminosity and the luminosity of high-energy neutrinos are directly coupled, the fraction of the optical energy that is emitted at the time the neutrino arrives should follow a uniform distribution. Within our search window of 1~yr after the optical peak, this fraction is 0.85, 0.87, and  0.66 for \bran, \tywin, and \lancel, respectively. A Kolmogorov-Smirnov test yields $p=0.08$ for the null hypothesis that these values are drawn from a uniform distribution. 

While the relatively late arrival time of the three neutrinos is not statistically significant, the effect is large enough to permit some speculation about possible origins for delayed neutrino emission in TDEs or flaring AGN. Below we suggest three possible explanations.

First, if the mass accretion rate is constant for about one year, the neutrino flux can also be expected to be constant over this period. A constant neutrino flux implies that a delayed neutrino detection is equally likely as a detection close to the optical peak. This is possible because the optical/UV emission of TDEs might not trace the accretion rate \citep{Piran15} and instead is emitted at the first shock of the stellar debris streams, which marks the onset of the debris circularization \citep{Bonnerot20_ISSI}, see \citet{Roth20_ISSI} for a review. A roughly constant accretion rate following the first year after disruption was used to explain the delayed neutrino detection of \bran, and this idea is supported by the radio \citep{Stein20} and X-ray observations of this event \citep{Mummery21}. 

Second, the mass accretion rate may not be constant, but delayed. 
%but decays with time following a power-law, $t^{-n}$. We have $1\lesssim{n}\le5/3$; the low end of $n$ is applicable for the disk-dominated phase \citep{Cannizzo95,vanVelzen18_FUV,Mummery20}, whereas the high end is for the standard TDE case \citep{Rees88,Phinney89}. 
The circularization timescale of the stellar debris can be estimated as 
\begin{eqnarray}
t_{\rm circ}\approx&& 0.55\,
\beta^{-3}
\left(\frac{\eta}{0.1}\right)^{-1} \nonumber \\ 
&&\left(\frac{M_{\rm bh}}{10^7\,M_\odot}\right)^{-7/6}
\,
\left(\frac{m_*}{1\,M_\odot}\right)^{8/25}\,
\,\,{\rm yr}
\label{eq:tcirc}
\end{eqnarray}
\citep{Hayasaki21}. Here $\eta$ represents how efficiently the kinetic energy at the stream-stream collision is dissipated; the most efficient ($\eta=1$) case corresponds to the result of \citet{2017MNRAS.464.2816B}. 
Because the inner accretion disk is formed after the circularization of the debris, it could take of order $t_{\rm circ}$ for the first neutrinos to be produced, which appears to match the observed time delay (Fig.~\ref{fig:lcs}). 

A caveat to Eq.~\ref{eq:tcirc} is a potential interaction of stellar debris stream with a pre-exisiting accretion disk, which can shorten the circularization timescale \citep{Chan19}. Another caveat is that the radiative efficiency of the first stream-stream shock might be too low to explain the prompt optical/UV emission \citep{Lu20}. Instead, the early-type optical/UV emission could be attributed to reprocessing of photons emitted from the accretion disk, implying the disk has already formed when the optical emission peaks \citep{Bonnerot20}. In this case, for PeV particle acceleration in the newly formed accretion disk, the debris circularization time is too short to explain the neutrino arrival delay.  

Finally, a delay between optical emission and the neutrino arrival time can be explained if particle acceleration happens in a jet or outflow and neutrinos are only produced when the resulting PeV protons collide with infrared photons of the dust echo \citep{Winter23}. A potential challenge to a jet-based particle acceleration mechanisms is discussed below in section~\ref{sec:puzzle}.

\subsection{Contribution to the high-energy neutrino flux}
About half of the neutrinos in the IceCube alerts dataset are expected to be background (i.e., atmospheric) events. Based on the ``signalness" probability \citep{ic_realtime} of the neutrinos in our sample, we expect about 16 astrophysical neutrinos in our sample. Hence three coincident events implies that at least $19_{-12}^{+22}\%$ (90\%CL) of the IceCube astrophysical neutrino alerts are explained by accretion flares in our sample. The fraction of the total astrophysical high-energy neutrino flux produced by the entire accretion flare population is larger than this estimate because our sample of flares is not complete. A factor $\approx 2$ increase could be expected to account for neutrino alerts from flares that are too distant to be detected by ZTF and NEOWISE \citep{Stein20}.

The use of dust echo properties to define the accretion flare population could provide another source of incompleteness. In this work, we use the echoes as tracers of energetic events in the accretion disk of massive black holes. A causal relation between the echo and the neutrino is not required for our analysis, but our search will, by construction, not identify neutrinos from TDEs in dust-free galaxies.

\subsection{Efficiency puzzle}\label{sec:puzzle}
A sizable contribution of accretion flares to the high-energy neutrino flux is remarkable because the energy we received from these flares is much lower compared to regular AGN. We can estimate the difference in fluence (energy received at Earth) of the two populations using the NEOWISE infrared observations. The sum of the infrared echo flux of the 63 accretion flares is 0.1 Jy, or a fluence of $10^{20.5}\,{\rm erg}\,{\rm cm}^{-2}$ for a one year duration of the flare. While the sum of the baseline infrared flux of the AGN detected by ZTF is 26 Jy, a fluence of $10^{23.3} \,{\rm erg}\,{\rm cm}^{-2}$ over the 3 year duration of our search (Fig.~\ref{fig:years}). Here we have only summed the infrared emission of AGN detected in the ZTF alert stream (due to their variability), if we add the contribution of the rest of the population, this estimate of the fluence could increase with another order of magnitude. 
%At radio wavelengths this difference in the received energy is even larger. 

Given that steady AGN outshine accretion flares by at least three orders of magnitude we reach a puzzling conclusion: in order to explain the observed PeV-scale neutrino associations, the accretion flares appear to be vastly more efficient at producing PeV neutrinos compared to the majority of AGN. Here we define efficiency as number of high-energy ($\sim$ PeV) particles relative to the electromagnetic energy output (i.e., $\eta_\nu$ in Eq.~\ref{eq:etanu}). 
%Since AGN outshine accretion flares by at least three orders of magnitude,
%Equal efficiency implies we should expect $\sim 10^{3}$ times more high-energy neutrinos from AGN compared to accretion flares. 
Since the contribution of accretion flares and normal AGN to the total neutrino flux have to add to 100\% (or less, if we also include other sources of neutrinos), equal efficiency would imply that accretion flares should produce at most $0.1\%$ of the total high-energy neutrino flux. This is clearly not consistent with the lower limit of 10\% based on the three accretion flares that are detected as potential neutrino sources.      

The fact that accretion flares are a cosmic minority presents a challenge for models of TDEs as neutrino sources that involve a relativistic jet \citep{fg08,Wang11,Winter:2020ptf} or a corona \citep{Murase20}. Since AGN also have these features, a similar efficiency of PeV-scale neutrino production can be expected for such models. %Different boundary conditions to allow more efficient particle acceleration in TDE-induced jets could be possible, but we deem this unlikely given that two of the three accretion flares with neutrino counterparts (\tywin and \lancel) occurred in AGN. 
Here we have formulated the efficiency puzzle in terms of high-energy neutrino production. But unless the optical depth for pion production is much lower for accretion flares compared to AGN, the same conclusions hold for the efficiency of high-energy particle acceleration.

The low black hole mass of the accretion flares coincident with neutrinos points to a potential solution for this efficiency puzzle. Both TDEs and extreme AGN flares are commonly observed to reach the Eddington limit \citep[e.g.,][]{Wevers17}, while the vast majority of AGN accrete an order of magnitude below the Eddington limit \citep[e.g.,][]{Kelly13}. Due to photon trapping, a high accretion rate increases the scale height of the accretion disk relative to the geometrically thin disk that operates at Eddington ratios of $\approx 10$\%. As the Eddington ratio approaches unity, we might expect a  {state change} of the accretion disk \citep{Abramowicz13}. In the context of TDEs, this {state change} is supported by observations: the X-ray spectra of typical AGN are hard and non-thermal, while TDE and extreme AGN flares have thermal and soft X-ray spectra \citep{Saxton20,Frederick20,Wevers20,Wevers21}. We can speculate that the accretion state that corresponds to high Eddington ratios enables more efficient particle acceleration. This possibility has been explored by \citet{Hayasaki19} for a magnetically arrested disk (MAD). The MAD accretion regime \citep{Narayan03} has also been employed by \citet{Scepi21} to explain the properties of a peculiar AGN flare (1ES\,1927+654: \citealt{Trakhtenbrot19}).

%Because geometrically thin disks are radiatively efficient and have a low temperature, the gas particles in these disks have a Maxwellian distribution, which disables the second-order Fermi acceleration. However, a super-Eddington accretion flow with a strong magnetic field ($\gtrsim 10^5$~G, see Fig.~\ref{fig:bfield} and Appendix~\ref{sec:acc}) could make PeV-scale particle acceleration possible \citep{Hayasaki19}.

Because the disk environment will absorb the gamma-ray emission produced in $\pi^0$ decay through the pair-creation process \citep{Hayasaki19,Murase20}, a disk-based particle accelerator is predicted to be dark above GeV energies, consistent with the upper limits on gamma-ray emission for the three accretion flares with neutrino counterparts (Table~\ref{tab:mm}).

In summary, the detection of three neutrinos from the energetically-subdominant population of accretion flares can be explained if the high-energy particle acceleration efficiency drastically increases towards the Eddington limit. 
%Outflows that are enabled by this strong magnetic field could explain why all three neutrino-emitting accretion flares in our sample are detected at radio wavelengths. 
This scenario might also provides an explanation for neutrino emission from NGC~1068, the most significant hotspot in the IceCube sky map at sub-PeV energies, detected at 4$\sigma$ post-trial \citep{Aartsen20,IceCube_NGC1068_22}. NGC~1068 is exceptional because it could be the nearest example of the small subset of AGN accreting near the Eddington limit \citep{Kawaguchi03,Lodato03}, similar to the accretion flares presented in this work.  
%If we narrow the scope of our signal hypothesis to include only accretion flares with evidence for super-Eddington accretion ($f_{\rm Edd}>0.5$; yielding 14 flares), our likelihood method yields a somewhat higher significance for the neutrino correlation (3.8$\sigma$). 
The small subset of persistent AGN that accrete close to the Eddington luminosity could  provide an important contribution to the potential correlation (detected at 2.6$\sigma$) between persistent AGN and IceCube neutrinos \citep{Abbasi22}.  \\*[10pt]

\section*{Acknowledgements}
%\begin{acknowledgments}
{ 
We acknowledge useful discussions and suggestions from  J.  Becerra Gonz\'alez, M. Kerr, W.~Lu, C.~Lunardini, K.~Murase, and W.~Winter. We thank the anonymous referee for the useful suggestions and comments.

Based on observations obtained with the Samuel Oschin Telescope 48-inch and the 60-inch Telescope at the Palomar Observatory as part of the Zwicky Transient Facility (ZTF) project. ZTF is supported by the National Science Foundation under Grant No. AST-1440341 and Grant No. AST-2034437 and a collaboration including Caltech, IPAC, the Weizmann Institute for Science, the Oskar Klein Center at Stockholm University, the University of Maryland, the University of Washington, Deutsches Elektronen-Synchrotron and Humboldt University, Los Alamos National Laboratories, the TANGO Consortium of Taiwan, the University of Wisconsin at Milwaukee, Trinity College Dublin, Lawrence Livermore National Laboratories, and IN2P3, France. Operations are conducted by COO, IPAC, and UW. SED Machine is based upon work supported by the National Science Foundation under Grant No. 1106171. 

This work is based on observations with the eROSITA telescope on board the SRG observatory. The SRG observatory was built by Roskosmos in the interests of the Russian Academy of Sciences represented by its Space Research Institute (IKI) in the framework of the Russian Federal Space Program, with the participation of the Deutsches Zentrum f\"ur Luft- und Raumfahrt (DLR). The SRG/eROSITA X-ray telescope was built by a consortium of German Institutes led by MPE, and supported by DLR. The SRG spacecraft was designed, built, launched, and is operated by the Lavochkin Association and its subcontractors. The science data are downlinked via the Deep Space Network Antennae in Bear Lakes, Ussurijsk, and Baykonur, funded by Roskosmos. The eROSITA data used in this work were processed using the eSASS software system developed by the German eROSITA consortium and proprietary data reduction and analysis software developed by the Russian eROSITA Consortium. 

This work includes data products from the Near-Earth Object Wide-field Infrared Survey Explorer (NEOWISE), which is a project of the Jet Propulsion Laboratory/California Institute of Technology. NEOWISE is funded by the National Aeronautics and Space Administration. The \textit{Fermi}-LAT Collaboration acknowledges support for LAT development, operation and data analysis from NASA and DOE (United States), CEA/Irfu and IN2P3/CNRS (France), ASI and INFN (Italy), MEXT, KEK, and JAXA (Japan), and the K.A.~Wallenberg Foundation, the Swedish Research Council and the National Space Board (Sweden). Science analysis support in the operations phase from INAF (Italy) and CNES (France) is also gratefully acknowledged. 

This work performed in part under DOE Contract DE-AC02-76SF00515. MG, PM and RS acknowledge the partial support of  this research by grant 21-12-00343 from the Russian Science Foundation. KH has been supported by the Basic Science Research Program through the National Research Foundation of Korea (NRF) funded by the Ministry of Education (2016R1A5A1013277 and 2020R1A2C1007219), and also financially supported during the research year of Chungbuk National University in 2021. 

The National Radio Astronomy Observatory is a facility of the National Science Foundation operated under cooperative agreement by Associated Universities, Inc. This research has made use of the CIRADA cutout service at URL cutouts.cirada.ca, operated by the Canadian Initiative for Radio Astronomy Data Analysis (CIRADA). CIRADA is funded by a grant from the Canada Foundation for Innovation 2017 Innovation Fund (Project 35999), as well as by the Provinces of Ontario, British Columbia, Alberta, Manitoba and Quebec, in collaboration with the National Research Council of Canada, the US National Radio Astronomy Observatory and Australia’s Commonwealth Scientific and Industrial Research Organisation.

AF received funding from the German Science Foundation DFG, within the Collaborative Research Center SFB1491 ``Cosmic Interacting Matters - From Source to Signal''. YY thanks the Heising–Simons Foundation for financial support. SR was supported by the Helmholtz Weizmann Research School on Multimessenger Astronomy, funded through the Initiative and Networking Fund of the Helmholtz Association, DESY, the Weizmann Institute, the Humboldt University of Berlin, and the University of Potsdam. ECK acknowledges support from the G.R.E.A.T research environment funded by {\em Vetenskapsr\aa det}, the Swedish Research Council, under project number 2016-06012, and support from The Wenner-Gren Foundations. MMK acknowledges generous support from the David and Lucille Packard Foundation. This work was supported by the GROWTH project funded by the National Science Foundation under Grant No 1545949.
}
%\end{acknowledgments}

\section*{Data Availability}
The data and software to reproduce the main results can be obtained via Zenodo \citep{HelloLancel}.

% \software{astropy \citep{2013A&A...558A..33A,2018AJ....156..123A}.           }

\bibliography{general_desk,extra}
\bibliographystyle{mnras}

\clearpage
%\appendix

% \section{Catalogs}
% In in Table~\ref{tab:IC} we give properties of the IceCube high-energy neutrino alerts and in Table~\ref{tab:ZW} we list the properties of the 63 accretion flares.

\begin{table*}
    \centering
    \begin{tabular}{lrrccl}
    \\
    \hline
    Event & Right Ascension & Declination & 90\% CL area  & Signalness & GCN ref. \\
    &  &  &(deg$^2$) \\
    \hline\hline
    IC180908A  & $144.58_{-1.55}^{+1.45}$ & $ -2.13_{-0.90}^{+1.20}$ & 6.3 & 0.34 & \cite{ic180908a}\\
IC181023A  & $270.18_{-2.00}^{+1.70}$ & $ -8.57_{-1.25}^{+1.30}$ & 9.3 & 0.28 & \cite{ic181023a}\\
IC190704A  & $161.85_{-4.33}^{+2.16}$ & $ 27.11_{-1.83}^{+1.81}$ & 21.0 & 0.49 & \cite{ic190704a}\\
IC190712A  & $ 76.46_{-6.83}^{+5.09}$ & $ 13.06_{-3.44}^{+4.48}$ & 92.1 & 0.30 & \cite{ic190712a}\\
IC190819A  & $148.80_{-3.24}^{+2.07}$ & $  1.38_{-0.75}^{+1.00}$ & 9.3 & 0.29 & \cite{ic190819a}\\
IC190922A  & $167.43_{-2.63}^{+3.40}$ & $-22.39_{-2.89}^{+2.88}$ & 32.2 & 0.51 & \cite{ic190922a}\\
IC191119A  & $230.10_{-6.48}^{+4.76}$ & $  3.17_{-2.09}^{+3.36}$ & 61.2 & 0.45 & \cite{ic191119a}\\
IC191122A  & $ 27.25_{-2.90}^{+1.70}$ & $ -0.04_{-1.49}^{+1.17}$ & 12.2 & 0.33 & \cite{ic191122a}\\
IC191204A  & $ 79.72_{-1.74}^{+3.20}$ & $  2.80_{-1.23}^{+1.12}$ & 11.6 & 0.33 & \cite{ic191204a}\\
IC191215A  & $285.87_{-3.19}^{+2.88}$ & $ 58.92_{-2.25}^{+1.85}$ & 12.8 & 0.47 & \cite{ic191215a}\\
IC191231A  & $ 46.36_{-3.47}^{+4.27}$ & $ 20.42_{-2.80}^{+2.11}$ & 35.5 & 0.46 & \cite{ic191231a}\\
IC200421A  & $ 87.93_{-2.83}^{+3.44}$ & $  8.23_{-1.84}^{+2.09}$ & 24.4 & 0.33 & \cite{ic200421a}\\
IC200425A  & $100.10_{-3.14}^{+4.67}$ & $ 53.57_{-1.60}^{+2.45}$ & 19.0 & 0.48 & \cite{ic200425a}\\
IC200523A  & $338.64_{-6.07}^{+10.77}$ & $  1.75_{-3.54}^{+1.84}$ & 90.5 & 0.25 & \cite{ic200523a}\\
IC200614A  & $ 33.84_{-6.39}^{+4.77}$ & $ 31.61_{-2.28}^{+2.75}$ & 47.9 & 0.42 & \cite{ic200614a}\\
IC200615A  & $142.95_{-1.45}^{+1.18}$ & $  3.66_{-1.06}^{+1.19}$ & 5.9 & 0.83 & \cite{ic200615a}\\
IC200806A  & $157.25_{-0.89}^{+1.21}$ & $ 47.75_{-0.64}^{+0.65}$ & 1.8 & 0.40 & \cite{ic200806a}\\
IC200911A  & $ 51.11_{-11.01}^{+4.42}$ & $ 38.11_{-1.99}^{+2.35}$ & 52.8 & 0.41 & \cite{ic200911a}\\
IC200921A  & $195.29_{-1.73}^{+2.35}$ & $ 26.24_{-1.77}^{+1.51}$ & 12.0 & 0.41 & \cite{ic200921a}\\
IC200926B  & $184.75_{-1.55}^{+3.64}$ & $ 32.93_{-0.91}^{+1.15}$ & 9.0 & 0.43 & \cite{ic200926b}\\
IC201014A  & $221.22_{-0.75}^{+1.00}$ & $ 14.44_{-0.46}^{+0.67}$ & 1.9 & 0.41 & \cite{ic201014a}\\
IC201114A  & $105.25_{-1.12}^{+1.28}$ & $  6.05_{-0.95}^{+0.95}$ & 4.5 & 0.56 & \cite{ic201114a}\\
IC201115A  & $195.12_{-1.49}^{+1.27}$ & $  1.38_{-1.11}^{+1.30}$ & 6.6 & 0.49 & \cite{ic201115a}\\
IC201120A  & $307.53_{-5.59}^{+5.34}$ & $ 40.77_{-2.80}^{+4.97}$ & 65.3 & 0.50 & \cite{ic201120a}\\
IC201221A  & $261.69_{-2.50}^{+2.29}$ & $ 41.81_{-1.20}^{+1.29}$ & 8.9 & 0.56 & \cite{ic201221a}\\
IC190503A  & $120.28_{-0.77}^{+0.57}$ & $  6.35_{-0.70}^{+0.76}$ & 1.9 & 0.36 & \cite{ic190503a}\\
IC190619A  & $343.26_{-2.63}^{+4.08}$ & $ 10.73_{-2.61}^{+1.51}$ & 27.1 & 0.55 & \cite{ic190619a}\\
IC190730A  & $225.79_{-1.43}^{+1.28}$ & $ 10.47_{-0.89}^{+1.14}$ & 5.4 & 0.67 & \cite{ic190730a}\\
IC190922B  & $  5.76_{-1.37}^{+1.19}$ & $ -1.57_{-0.82}^{+0.93}$ & 4.5 & 0.51 & \cite{ic190922b}\\
IC191001A  & $314.08_{-2.26}^{+6.56}$ & $ 12.94_{-1.47}^{+1.50}$ & 25.5 & 0.59 & \cite{ic191001a}\\
IC200109A  & $164.49_{-4.19}^{+4.94}$ & $ 11.87_{-1.36}^{+1.16}$ & 22.5 & 0.77 & \cite{ic200109a}\\
IC200117A  & $116.24_{-1.24}^{+0.71}$ & $ 29.14_{-0.78}^{+0.90}$ & 2.9 & 0.38 & \cite{ic200117a}\\
IC200512A  & $295.18_{-2.26}^{+1.72}$ & $ 15.79_{-1.29}^{+1.26}$ & 9.8 & 0.32 & \cite{ic200512a}\\
IC200530A  & $255.37_{-2.56}^{+2.48}$ & $ 26.61_{-3.28}^{+2.33}$ & 25.2 & 0.59 & \cite{ic200530a}\\
IC200620A  & $162.11_{-0.95}^{+0.64}$ & $ 11.95_{-0.48}^{+0.63}$ & 1.7 & 0.32 & \cite{ic200620a}\\
IC200916A  & $109.78_{-1.44}^{+1.08}$ & $ 14.36_{-0.85}^{+0.88}$ & 4.2 & 0.32 & \cite{ic200916a}\\
IC200926A  & $ 96.40_{-0.55}^{+0.73}$ & $ -4.33_{-0.76}^{+0.61}$ & 1.7 & 0.43 & \cite{ic200926a}\\
IC200929A  & $ 29.53_{-0.53}^{+0.53}$ & $  3.47_{-0.35}^{+0.71}$ & 1.1 & 0.47 & \cite{ic200929a}\\
IC201007A  & $265.17_{-0.52}^{+0.52}$ & $  5.34_{-0.23}^{+0.32}$ & 0.6 & 0.89 & \cite{ic201007a}\\
IC201021A  & $260.82_{-1.68}^{+1.73}$ & $ 14.55_{-0.74}^{+1.35}$ & 6.9 & 0.30 & \cite{ic201021a}\\
IC201130A  & $ 30.54_{-1.31}^{+1.13}$ & $-12.10_{-1.13}^{+1.15}$ & 5.4 & 0.15 & \cite{ic201130a}\\
IC201209A  & $  6.86_{-1.22}^{+1.02}$ & $ -9.25_{-1.14}^{+0.99}$ & 4.7 & 0.19 & \cite{ic201209a}\\
IC201222A  & $206.37_{-0.80}^{+0.90}$ & $ 13.44_{-0.38}^{+0.55}$ & 1.5 & 0.53 & \cite{ic201222a}\\

    \end{tabular}
    \\[7pt]
    \caption{IceCube neutrino alerts used in this study.}
    \label{tab:IC}
\end{table*}

\clearpage
\newpage
\onecolumn 
{
%\fontsize{9}{11}\selectfont
%\begin{table}[]
\begin{longtable}{lccccccccc}
    %\begin{tabular}{lcccccccc}
    \hline
    ZTF name & $t_{\rm peak}^{a}$ & rise$^{b}$ & fade$^{b}$ & $\Delta F_{\rm IR}/F_{\rm rms}$  & $\Delta F_{\rm IR}$ & $z$ & $M_{\rm BH}$ & \multirow{2}*{$\frac{P({\rm AGN})}{P(\rm TDE)}^{c}$}  & spectro. \\
    & (MJD) & (days) & (days) &  & (mJy) & &($\log_{10} M_\odot$) & & class\\
    \hline\hline
    AT2019dsg&58620.2&32.1&81.9&92.2&$1.58\pm0.02$&0.0512&6.74 (C21)&0.000&TDE\\
AT2019fdr&58672.5&30.8&336.6&39.2&$0.71\pm0.07$&0.2666&7.10 (F20)&0.000&TDE?\\
AT2019aalc&58658.2&49.0&167.7&15.7&$11.13\pm0.10$&0.0356&7.23 (L19)&0.146&TDE?\\
\hline
AT2018dyk&58261.4&60.0&342.0&23.8&$1.41\pm0.03$&0.0367&5.50 (F19)&0.000&TDE?\\
AT2019aame&58363.2&--&138.0&12.3&$0.18\pm0.01$&--&--&0.540&--\\
AT2018lzs&58378.2&134.0&15.5&$\phantom{1}$3.3&$0.03\pm0.01$&--&--&3.651&--\\
AT2021aetz&58390.3&8.6&--&47.5&$0.81\pm0.01$&0.0879&6.21 (T13)&0.000&TDE?\\
AT2018ige&58432.5&--&67.9&65.8&$0.31\pm0.01$&--&--&0.000&--\\
AT2021aeud&58448.3&137.6&282.7&$\phantom{1}$6.2&$0.15\pm0.01$&--&--&3.937&--\\
AT2018iql&58449.4&17.9&41.0&30.1&$0.48\pm0.01$&--&--&0.000&--\\
AT2018jut&58449.6&--&--&$\phantom{1}$5.0&$0.12\pm0.01$&--&--&5.424&--\\
AT2021aeue&58475.1&48.5&48.8&$\phantom{1}$4.9&$0.11\pm0.01$&--&--&5.552&--\\
AT2019aamf&58506.4&80.0&141.2&$\phantom{1}$6.6&$0.19\pm0.01$&--&--&3.395&--\\
AT2018kox&58510.2&26.9&167.6&$\phantom{1}$5.6&$0.33\pm0.01$&0.096&--&4.668&TDE?\\
AT2018lhv&58513.5&17.5&116.1&32.3&$0.35\pm0.01$&--&--&0.000&--\\
AT2019avd&58534.3&14.9&52.9&67.5&$1.87\pm0.05$&0.0296&6.10 (F20)&0.000&TDE?\\
AT2016eix&58539.4&104.2&43.1&$\phantom{1}$6.9&$0.16\pm0.01$&--&--&2.922&--\\
AT2019aamg&58540.5&--&93.7&$\phantom{1}$8.3&$0.14\pm0.00$&--&--&1.612&--\\
AT2018lcp&58547.2&95.7&297.4&12.7&$0.16\pm0.01$&0.06&--&0.482&TDE?\\
AT2021aeuf&58556.4&17.0&137.5&15.6&$0.18\pm0.01$&--&--&0.155&--\\
AT2020aezy&58558.4&82.5&359.8&$\phantom{1}$4.8&$0.06\pm0.01$&--&--&5.623&--\\
AT2019cle&58568.4&9.2&54.8&19.4&$0.25\pm0.01$&--&--&0.012&--\\
AT2019aamh&58582.5&--&154.5&$\phantom{1}$7.7&$0.40\pm0.01$&--&--&2.095&--\\
AT2019dll&58605.2&29.6&--&$\phantom{1}$6.8&$0.26\pm0.01$&0.101&7.48 (T13)&3.109&TDE?\\
AT2019gur&58607.5&--&234.5&38.6&$0.34\pm0.01$&--&--&0.000&--\\
AT2018lof&58608.2&70.9&370.4&$\phantom{1}$4.1&$0.15\pm0.01$&0.302&8.98 (L19)&6.078&AGN\\
AT2019dqv&58628.2&67.0&475.0&40.4&$1.77\pm0.02$&0.0816&6.67 (L19)&0.000&TDE?\\
AT2019cyq&58637.2&49.5&95.0&31.8&$0.54\pm0.01$&0.262&7.56 (L19)&0.000&TDE?\\
AT2021aeug&58641.2&48.6&68.8&$\phantom{1}$4.6&$0.11\pm0.01$&--&--&5.900&--\\
AT2019ihv&58646.5&--&19.3&$\phantom{1}$8.7&$0.29\pm0.01$&0.1602&--&1.395&--\\
AT2019dzh&58651.2&51.4&348.9&$\phantom{1}$6.4&$0.15\pm0.01$&0.314&--&3.659&--\\
AT2019kqu&58652.2&95.6&315.6&$\phantom{1}$6.1&$0.27\pm0.01$&0.174&7.53 (L19)&3.996&TDE?\\
AT2019hbh&58652.3&19.8&106.7&$\phantom{1}$8.4&$0.75\pm0.01$&--&--&1.556&--\\
AT2020aezz&58677.3&99.8&280.6&$\phantom{1}$5.8&$0.18\pm0.01$&--&--&4.405&--\\
AT2020afaa&58678.2&83.8&330.3&$\phantom{1}$7.0&$0.13\pm0.02$&--&--&2.830&--\\
AT2019idm&58682.2&51.7&--&25.2&$0.41\pm0.02$&0.0544&6.64 (T13)&0.000&TDE?\\
AT2019ihu&58709.5&79.7&110.9&$\phantom{1}$6.2&$0.25\pm0.02$&0.27&8.90 (L19)&3.843&AGN\\
AT2019meh&58713.1&23.2&127.9&29.7&$1.99\pm0.04$&0.0935&7.06 (V23)&0.000&TDE?\\
AT2020afab&58717.2&47.5&127.2&$\phantom{1}$5.0&$0.22\pm0.01$&0.2875&6.49 (V23)&5.432&TDE?\\
AT2019aami&58717.4&--&152.0&31.8&$0.41\pm0.01$&--&--&0.000&--\\
AT2019nna&58717.4&34.4&87.1&27.0&$0.36\pm0.01$&--&--&0.000&--\\
AT2019nni&58732.2&29.1&109.5&$\phantom{1}$4.9&$0.36\pm0.01$&0.137&--&5.489&--\\
AT2021aeuk&58733.1&43.4&238.1&$\phantom{1}$7.3&$0.33\pm0.01$&0.235&--&2.464&--\\
AT2019hdy&58749.5&--&87.9&$\phantom{1}$4.0&$0.20\pm0.01$&0.442&--&6.000&--\\
AT2019pev&58750.1&13.1&54.7&$\phantom{1}$7.4&$0.19\pm0.01$&0.097&6.40 (F20)&2.364&TDE?\\
AT2019qiz&58753.1&11.9&34.7&44.4&$2.14\pm0.04$&0.01499&6.19 (N20)&0.000&TDE\\
AT2019brs&58758.1&110.9&483.2&$\phantom{1}$9.6&$0.78\pm0.01$&0.3736&7.20 (F20)&1.060&TDE?\\
AT2020afac&58758.3&68.0&95.0&10.8&$0.16\pm0.01$&--&--&0.802&--\\
AT2019wrd&58764.3&82.6&--&$\phantom{1}$7.6&$0.15\pm0.01$&--&--&2.197&--\\
AT2021aeuh&58789.5&--&--&$\phantom{1}$3.9&$0.28\pm0.01$&0.0834&7.39 (L19)&5.867&TDE?\\
AT2019msq&58791.2&137.9&--&$\phantom{1}$6.4&$0.16\pm0.01$&--&--&3.586&--\\
AT2019qpt&58798.3&44.5&135.1&13.8&$0.24\pm0.01$&0.242&6.97 (L19)&0.341&TDE?\\
AT2020afad&58802.2&77.1&125.5&$\phantom{1}$3.7&$0.08\pm0.01$&--&--&5.216&--\\
AT2019mss&58811.6&35.2&209.7&20.8&$0.38\pm0.01$&--&--&0.004&--\\
AT2019thh&58851.1&104.7&433.8&72.2&$1.75\pm0.02$&0.0506&--&0.000&TDE?\\
AT2021aeui&58860.3&--&60.0&$\phantom{1}$6.2&$0.49\pm0.02$&--&--&3.899&--\\
AT2020afae&58867.2&30.2&42.0&$\phantom{1}$5.2&$0.35\pm0.02$&--&--&5.200&--\\
AT2020mw&58867.3&12.6&32.1&$\phantom{1}$6.8&$0.12\pm0.01$&--&--&3.061&--\\
AT2020iq&58878.1&22.7&71.5&24.5&$0.45\pm0.01$&0.096&6.37 (V23)&0.000&TDE?\\
AT2019xgg&58891.2&81.6&446.2&$\phantom{1}$4.4&$0.13\pm0.01$&--&--&6.003&--\\
AT2020atq&58903.2&39.3&205.5&20.8&$0.66\pm0.01$&--&--&0.003&--\\
AT2021aeuj&58974.2&79.0&233.5&18.1&$0.20\pm0.01$&0.695&--&0.032&--\\
AT2020hle&58978.3&32.7&62.9&21.0&$0.33\pm0.01$&0.103&6.40 (F20)&0.003&TDE?\\
\\[7pt]
    \caption{Optical and infrared properties of the accretion flares used in this work.\\
    Notes --- The first three entries of this table list the events coincident with an IceCube neutrino alert. $^{a}$The column $t_{\rm peak}$ lists the time of maximum light of the ZTF light curve. $^{b}$The rise and fade columns list the $e$-folding time. $^{c}$The ratio of the AGN and TDE probability is based on the strength of the dust echo ($\Delta F_{\rm IR}/F_{\rm rms}$, see Fig.~\ref{fig:h_strength}). The references listed behind the black hole mass estimates give the origin of the optical spectrum that was used: T13 \citep{Thomas13}, L19 \citep{Liu19_SDSS}, F19 \citep{Frederick19}, F20 \citep{Frederick20}, N20 \citep{Nicholl20},  C21 \citep{Cannizzaro21}, V23 (this work). }
    \label{tab:ZW}
%\end{table}
\end{longtable}
}

%\bsp	% typesetting comment
\label{lastpage}
\end{document}